\documentclass[preprint2]{aastex}
\usepackage{picture}
\usepackage{graphicx}
\usepackage{color}

\newcommand{\gtabouteq}{\,\hbox{\raise 0.5 ex \hbox{$>$}\kern-.77em 
                    \lower 0.5 ex \hbox{$\sim$}$\,$}}       
\newcommand{\ltabouteq}{\,\hbox{\raise 0.5 ex \hbox{$<$}\kern-.77em 
                     \lower 0.5 ex \hbox{$\sim$}$\,$}}

\slugcomment{To appear in the Astronomical Journal}

\shorttitle{CHANG-ES: III}
\shortauthors{Irwin et al.}

\begin{document}


\title{CHANG-ES III: UGC10288 -- An Edge-on Galaxy with a Background Double-lobed Radio Source}


\author{Judith Irwin\altaffilmark{1}, 
Marita Krause\altaffilmark{2},
Jayanne English\altaffilmark{3},
Rainer Beck\altaffilmark{4},
Eric Murphy\altaffilmark{5}, 
Theresa Wiegert\altaffilmark{6},
George Heald\altaffilmark{7}, 
Rene Walterbos\altaffilmark{8},
Richard J. Rand\altaffilmark{9},
}
\and
\author{Troy Porter\altaffilmark{10}}


\altaffiltext{1}{Dept. of Physics, Engineering Physics \& Astronomy, 
Queen's University, Kingston, ON, Canada, K7L 3N6, {\tt irwin@astro.queensu.ca}.}
\altaffiltext{2}{Max-Planck-Institut f{\"u}r Radioastronomie,  Auf dem H{\"u}gel 69,
53121, Bonn, Germany,
{\tt mkrause@mpifr-bonn.mpg.de}.} 
\altaffiltext{3}{Department of Physics and Astronomy, 
University of Manitoba, Winnipeg, Manitoba, Canada, R3T 2N2,
{\tt jayanne\_english@umanitoba.ca}.}
\altaffiltext{4}{Max-Planck-Institut f{\"u}r Radioastronomie, Auf dem H{\"u}gel 69,
53121, Bonn, Germany,
{\tt rbeck@mpifr-bonn.mpg.de}.}
\altaffiltext{5}{Observatories of the Carnegie 
Institution for Science, 813 Santa Barbara Street, Pasadena, CA, 91101,
USA,  {\tt 
emurphy@obs.carnegiescience.edu}.} 
\altaffiltext{6}{Dept. of Physics, Engineering Physics \& Astronomy,
Queen's University, Kingston, ON, Canada, K7L 2T3, 
{\tt twiegert@astro.queensu.ca}.} 
\altaffiltext{7}{Netherlands Institute for Radio Astronomy (ASTRON), 
Postbus 2, 7990 AA, Dwingeloo, The Netherlands,
{\tt heald@astron.nl}.}
\altaffiltext{8}{Dept. of Astronomy, New Mexico State University, 
PO Box 30001, MSC 4500, Las Cruces, NM 88003, USA, {\tt rwalterb@nmsu.edu}.}
\altaffiltext{9}{Dept. of Physics and Astronomy, University of New Mexico, 
800 Yale Boulevard, NE, Albuquerque, NM, 87131, USA, {\tt rjr@phys.unm.edu}.} 
\altaffiltext{10}{Hansen Experimental Physics Laboratory, Stanford University, 
452 Lomita Mall, Stanford, CA, 94305, USA, {\tt tporter@stanford.edu}.}



\begin{abstract}
This 3rd paper in the CHANG-ES series shows the first results from our regular
data taken with the Karl G. Jansky Very Large Array (JVLA).  The
edge-on galaxy, UGC~10288, has been observed in the B, C, and D configurations at 
L-band (1.5 GHz) and in the C and D configurations at C-band (6 GHz) in all polarization
products.  We show the first spatially resolved images of this galaxy in these bands, 
the first polarization images, and the first composed image at an intermediate frequency (4.1 GHz) which has
been formed from a combination of all data sets.

A surprising new result is the presence of a strong, polarized, double-lobed extragalactic radio source
({\it CHANG-ES A})
almost immediately behind the galaxy and perpendicular to its disk. The core
of {\it CHANG-ES A} has an optical counterpart 
(SDSS J161423.28-001211.8)
at a photometric
redshift of $z_{phot}\,=\,0.39$; the southern radio lobe is behind the disk of UGC~10288 and 
the northern lobe is behind the halo region.  
   This background `probe' has allowed us
to do a preliminary Faraday Rotation analysis of the foreground galaxy, putting
limits on the regular magnetic field and electron density in the halo of UGC~10288 in regions in
which there is no direct detection of a radio continuum halo.

We have revised the flux densities of the two sources individually as well as the star formation
rate (SFR) for UGC~10288.  The SFR is low (0.4 to 0.5 M$_\odot$ yr$^{-1}$) and the galaxy has a
high thermal fraction (44\% at 6 GHz), as estimated using both the thermal and non-thermal SFR
calibrations of \cite{mur11}.  UGC~10288 would have fallen well below the CHANG-ES flux density
cutoff, had it been considered without the brighter contribution of the background source.

UGC~10288 shows discrete high-latitude radio continuum features, but it does not have a
{\it global} radio continuum halo (exponential scales heights are typically
$\approx$ 1 kpc averaged over regions with and without extensions). 
 One prominent feature appears to form a large arc to the north of the galaxy on
its east side, extending to 3.5 kpc above the plane.  The total minimum magnetic field strength
at a sample position in
the arc is $\sim$ 10 $\mu$G.  Thus, this galaxy still appears to be able to form
 substantial high latitude, localized features in spite of its relatively low SFR.

\end{abstract}





\section{Introduction}
\label{sec:introduction}

This is the 3rd paper in the series, ``Continuum Halos in Nearby Galaxies --
an EVLA Survey'' (CHANG-ES) whose goals are to map a sample of 35 edge-on galaxies
at L-band (1.5 GHz) and C-band (6.0 GHz) 
in all four Stokes parameters, in a variety of configurations of the
Expanded Very Large Array (EVLA).  The EVLA is now known as the Karl G. Jansky
Very Large Array (JVLA) of the National Radio Astronomy Observatory\footnote{The National 
Radio Astronomy Observatory is a facility of the National Science Foundation 
operated under cooperative agreement by Associated Universities, Inc.}
and we will refer to it as the JVLA from now on.  A complete
description of the galaxy sample and goals of the survey can be found in
\cite{irw12a} (hereafter, Paper 1) and first results based on early test data of NGC~4631 can be
found in \cite{irw12b} (hereafter, Paper 2).

UGC~10288 (Table~\ref{table:prop-ugc10288})
is a member of the Lyon Group of Galaxies 404 (LGG-404) \citep{gar93} 
which contains only two other, widely separated
members: UGC~10290, 1.0 deg (0.61 Mpc, assuming the distance of UGC~10288) to the north, and 
NGC~6070, 1.4 deg (0.86 Mpc) to the north-west.  UGC~10288, itself, is edge-on
(inclination of 90 deg) and is listed in the Revised Flat Galaxy Catalog \citep{kar99}.

UGC~10288 has not previously been observed extensively, but
 it has
been a target in a few searches for high latitude emission. 
\cite{ran96} detected a few H$\alpha$ spurs extending to 1 kpc below the 
galaxy's disk at an emission measure 
level of 40 pc cm$^{-6}$. 
Follow-up single-slit optical spectra at two positions \citep{col01} then revealed emission lines to a vertical
z height\footnote{The coordinate, z, is taken to be the perpendicular distance from
the galaxy's plane.} of z = 3 kpc
and showed that the line ratios were best described by composite
photoionization/shock models.
An H$\alpha$ observation by 
\cite{ros00} was too low in signal-to-noise (S/N) to detect more than the brightest disk emission.
\cite{alt00} claim a dust absorption exponential scale height that is unresolved
at 1.2 arcsec resolution (198 pc). 
Of these references, only \cite{ran96} has published an original image
of the galaxy.

Aside from the 1.4 GHz National Radio Astronomy Observatory VLA Sky Survey (NVSS), whose rms noise is  
450 $\mu$Jy beam$^{-1}$ in a 45 arcsec beam \citep{con98},
the only previous radio continuum observation of UGC~10288 was made by \cite{hum91} at 5.0 GHz
to an rms limit of 90 $\mu$Jy beam$^{-1}$ with a beam of 14.5$^{\prime\prime}$.  The 
latter image shows essentially
a point source with a slight elongation to the north which has been interpreted as evidence for
a radio halo \citep{ros03}.   Because of these new JVLA observations, however, we now 
know that this elongation is due to a strong background 
double-lobed extragalactic
radio source (EGRS).  Our CHANG-ES observations achieve a best rms of 3 $\mu$Jy beam$^{-1}$ for a 
single array/frequency band combination
(C-array, C-band), and provide a range of spatial resolution from
3.0 arcsec to 37 arcsec,
significantly improving on these previous
data.  


\section{Observations and Data Reductions}
\label{sec:obs_red}

Observations and data reductions of UGC~10288 were similar to those described for
the test data in Paper 2.  However, there were also some differences\footnote{For example,
our correlator setup is different to avoid some strong, persistent interference, our observations
were staggered in time, and we have made delay corrections which were not
standardly available previously.}
and, since
this is the first paper describing our `regular' (as opposed to `test') observations, we describe each step
fully below.  Note that these data reductions are consistent with current standard continuum tutorials\footnote{See
http://casaguides.nrao.edu/index.php?title=EVLA\_Tutorials.}, unless otherwise indicated.

\subsection{Observations of UGC~10288}
\label{sec:observations}

Observations were carried 
out using the JVLA during its commissioning period {in
L-band (B, C, and D arrays) and at C-band (C and D arrays)}.
A log of the observational set-up for each JVLA configuration and each frequency band  
is given in Table~\ref{table:obs}.  A single array and frequency combination will be referred to
as an `observation' in the following.  As can be seen, the same 
calibrators were used for each observation, and the same correlator set-up was used.  
We introduced a gap in the
frequency band coverage at L-band
(from 1.503 to 1.647 GHz) because of persistent strong interference in that frequency range.

The observations were carried out in a standard fashion with scans of the source 
flanked by scans on a complex gain calibrator (J1557--0001)
that was near it in the sky. We will refer to this source as the `secondary calibrator'.
In addition, a scan was made on a bright calibrator of known flux density
and brightness distribution
(3C~286), i.e. the `primary calibrator' (this source is also used for the bandpass 
and polarization angle calibration).
Finally, a scan was also carried out on
a calibrator that was known to have negligible polarization (J1407+2827 = OQ~208), the
latter to determine the polarization leakage (see Sect.~\ref{sec:P_cal_imag}).  
The longest consecutive scan
on the galaxy at any array, between scans of the secondary calibrator, was 25 minutes.
Scans of 3C286, J1557--0001, and J1407+2827 
were typically, 7.5 to 11 minutes, 4.5 minutes, and 8 to 11 minutes,  respectively.
For each observation, the galaxy was observed within a scheduling block (SB) that included other
galaxies.  This allowed us to observe any given galaxy near the beginning of the SB and then again near the
end, so that the broadest possible uv\footnote{In this paper, we use `uv' to describe the plane
within which the antennas are 
distributed on the ground, and `UV' to specify the ultraviolet part of the spectrum.}
coverage could be achieved.  
For UGC~10288, the gaps in time varied with array and frequency, ranging from about 1.5 to 4.8 hours.

\subsection{Total Intensity Calibration}
\label{sec:I_calibration}

Data were reduced using the Common Astronomy Software
Applications (CASA) package\footnote{\tt http://casa.nrao.edu},
versions 3.3 and 3.4, and each observation was reduced in an identical fashion,
 as described below, unless otherwise indicated.
Note that JVLA antennas detect right (R) and left (L) circularly polarized radiation.
In the following description, R and L are treated separately in any
antenna-based calibration and the parallel hands (RR and LL) are treated separately in any
baseline-based calibration.  Calibration of the cross-hands (RL and LR) is discussed in 
Sect.~\ref{sec:P_cal_imag}

A copy of the data set was first made and then Hanning smoothed so as to more easily detect bad data.  An initial
pass of flagging then took place to remove the worst of the radio frequency interference (RFI)\footnote{All flagging
has been done manually.}. 
An antenna position correction table was then produced, if needed, based
on known positions as posted by NRAO\footnote{\tt http://www.vla.nrao.edu/astro/archive/baselines/.}. 
 An antenna-based 
delay correction table
 was then formed using one-minute of the scan on the primary calibrator, 
during which the antenna position correction table was applied on the fly.  Delay errors produce
phase slopes as a function of frequency; we found delay corrections of 
6 ns or less, as determined with respect to a reference antenna in the same polarization (R or L).
The same RFI flags determined in the first pass of the Hanning smoothed data
as well as the antenna position correction and delay correction tables
 were then applied to the original non-Hanning smoothed data. 
This ordering of steps (which is not described in standard CASA tutorials) 
was needed because delay corrections must
be applied to non-Hanning-smoothed data\footnote{If there were antennas with large delays, then delay errors, which are a function
of frequency, lead to
loss of amplitude which in turn introduces non-closing errors. As a precaution, we ensured that delay corrections were carried
out on data with the full spectral resolution.}, 
whereas locating RFI is most effectively done on Hanning-smoothed data.
 The corrected data were then
split off into a separate data set and Hanning-smoothed for further processing.  

Another round of RFI flagging was then carried out until there were no obvious 
amplitude departures from typical values.
In addition, each observing band contains contiguous spectral windows (Table~\ref{table:obs}),
 each with its own bandpass response.
Since the bandpass response in each spectral window declines strongly at the upper
and lower edges, we flagged 5 channels at both the upper and lower ends of each spectral window. JVLA observing
parameters did not allow us to overlap the spectral windows without losing significant
total bandwidth\footnote{The exception is at the center of the entire
observing band; at C-band, the upper and lower halves of the band overlap slightly.}
so that, after flagging, there are gaps of 10 channels between each spectral window.
The flux density of the primary calibrator
 was then set for each channel in the band using a known 
model\footnote{The Perley-Butler, 2010 model, available in the CASA routine, {\it setjy}, was used.} for Stokes I.

An initial calibration table of phase as a function of time for the primary calibrator
was made for a small range of channels at the center of each
spectral window (where the bandpass response is reasonably flat).  A bandpass calibration table was
then made using the same calibrator, applying the previous phase calibration table on the fly.
Application of the
initial phase calibration table ensures that no decorrelation occurs when vector averaging the data to determine
bandpass solutions.
Bandpasses were determine for each antenna and
each spectral window.  The initial phase calibration table is then no longer used.

Phase and amplitude calibration tables were then determined separately 
(first phase and then amplitude) for the primary calibrator, the secondary calibrator, and
the polarization leakage calibrator.
The bandpass calibration table was applied on the fly during these steps and the phase calibration table
was also applied on the fly when the amplitude calibration table was formed.  Phase 
solutions were determined for each integration
time for the primary calibrator, and for each scan for the
secondary and polarization leakage calibrators. Amplitude solutions
were determined per scan for all calibrators.
Flux densities for the secondary and polarization leakage calibrators were then found from the
known primary calibrator values on a per spectral window basis. 
Results for a representative frequency near the center of
each band are given in Table~\ref{table:obs}.  The results at L-band over the various independent observations
 agree to 1\% for J1407+2827
and 3\% for J1557--0001.  At C-band, the agreement is to within 0.2\% for J1407+2827 and 1\% for J1557--0001.

All calibration tables were then applied to the data 
(calibrators plus galaxy) to form a calibrated data set in total intensity, I.
For the galaxy, the calibration solutions determined from
the secondary calibrator were applied, interpolated linearly with time.  At this point,
low level RFI, which was not evident in the uncalibrated data, became much more obvious.  
Therefore another round of
flagging ensued and the above steps were all repeated, with new tables made and a new calibration applied.  
The flux
density data
for the calibrators presented in
Table~\ref{table:obs} refer to the final results.

\subsection{Total Intensity Imaging and Self-Calibration}
\label{sec:I_imaging}
Total intensity images were formed for each observation by first Fourier Transforming the uv data
into the sky plane (forming the `dirty map')
and then deconvolving the `dirty beam' (i.e. the point-spread function) from
the dirty map
using the Clark CLEAN algorithm \citep{cla80}.  Further details regarding our imaging of wide-band data can be found
in Paper 2 and the parameters used to make our maps are in Table~\ref{table:map-ugc10288}.


\begin{figure}[h]
   \centering
   \includegraphics*[width=3.2in]{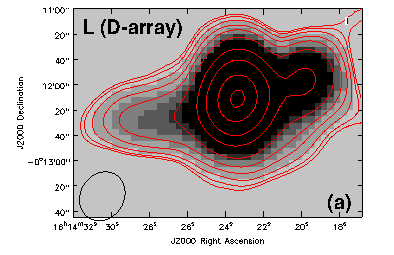}
   \hspace{0in}
   \includegraphics*[width=3.2in]{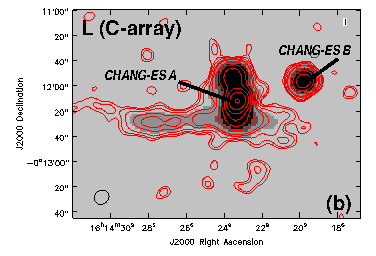}
   \hspace{0in}
   \includegraphics*[width=3.2in]{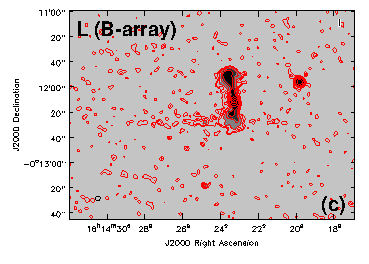}
\caption{L-band images of UGC~10288 (both greyscale and contours).  
The beam size is shown as a black ellipse at
the bottom left and  corresponding map parameters are given in 
Table~\ref{table:map-ugc10288}.   {\it (a)} D-array map; contours are at
84 (2$\sigma$), 120, 200, 350, 600, 1200, 3000, 6000, 12000, and 18000 $\mu$Jy beam$^{-1}$.
{\it (b)} C-array map; contours are at 44  (2$\sigma$), 60, 90, 150, 200, 400, 800, 1600, 3000, 6000, 
and 12000 $\mu$Jy beam$^{-1}$.  The background source is labelled.
{\it (c)} B-array map; contours are at
28 (2$\sigma$), 50, 100, 200, 400, 1200, 3600, and 8000 $\mu$Jy beam$^{-1}$.
}
\label{fig:Lband}
\end{figure}

\begin{figure}[h]
   \centering
   \includegraphics*[width=3.3in]{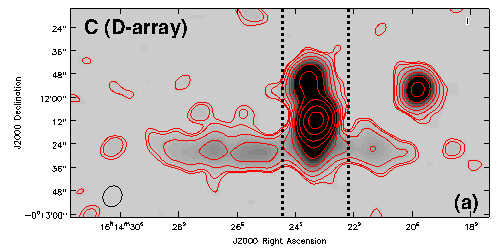}
   \includegraphics*[width=3.3in]{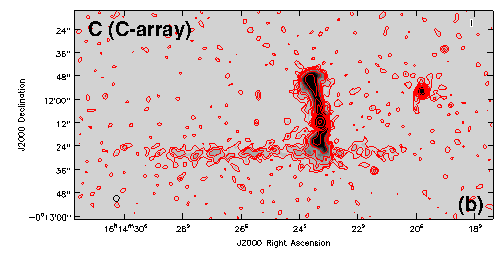}
\caption{C-band images of UGC~10288 (both greyscale and contours).  
The beam size is shown as a black ellipse at
the bottom left and  corresponding map parameters are given in 
Table~\ref{table:map-ugc10288}.   {\it (a)} D-array map; contours are at
14 (2$\sigma$), 20, 40, 80, 150, 300, 600, 1200, 3000, and 6000 $\mu$Jy beam$^{-1}$.
Dashed lines denote the region used to estimate the fluxes of 
Table~\ref{table:derived-ugc10288}. 
{\it (b)} C-array map; contours are at
6.2 (2$\sigma$), 12, 20, 30, 60, 120, 200, 400, 1600, and 6000 $\mu$Jy beam$^{-1}$.
}
\label{fig:Cband}
\end{figure}

In brief, we used the wide field imaging algorithm \citep{cor08b} and
 multi-scale, multi-frequency synthesis (ms-mfs) during imaging
\citep{cor08a,rau11}. Wide field imaging takes into account the fact that all antennas may not
be co-planar.  
The multi-scale clean assumes that the emission can be 
represented by a variety of spatial scales, rather than only point sources as has been largely
done in the past. 
Multi-frequency synthesis makes use of
the fact that, at different frequencies within the band, the uv spacings (measured in k$\lambda$) are also different and therefore the
uv plane is more filled in than would be the case for  monochromatic observations.
We also fit the emission across the band in each pixel with a spectrum  \citep{sau94} of the form,
$I_\nu\,\propto\,\nu^\alpha$, which, in practice, is expanded as a Taylor series with 2 terms (see Paper 2).
At the end of the cleaning process, the image is constructed by 
reconvolving the clean components with a Gaussian of the same full-width-half-maximum
(FWHM) as the dirty beam.

Each data set 
 was then self-calibrated \citep[e.g.][]{pea84}, with the exception of D-array L-band for which self-calibration attempts 
did not improve the map.  The number and type of self-calibration iterations is given 
in Table~\ref{table:map-ugc10288}.

The resulting total intensity images are seen in Figs.~\ref{fig:Lband} and \ref{fig:Cband} for the L-band and C-band data, respectively.  
We additionally
formed images (not shown)
 at spatial resolutions intermediate between those shown in these figures (e.g. by tapering the uv plane response).
The images of Figs.~\ref{fig:Lband} and \ref{fig:Cband} 
are shown without a primary beam correction, but any flux density (or other numerical) measurements are made on primary beam-corrected images.
The primary beam correction was carried out as described in Paper 2, except that the beam has now
been constructed with frequency weighting that matches the frequency weighting of the data (for
example, the weighting that necessarily results from
flagging)\footnote{Available in CASA 4.0.0 and higher.}

\subsection{Polarization Calibration and Imaging}
\label{sec:P_cal_imag}

Polarization calibration proceeded in the same fashion for all data sets.
The cross correlation data (RL and LR) were first inspected separately and 
additional flagging was carried out as needed.  A known model for Stokes Q and U for the primary
calibrator was introduced, based on formulae given in Paper 2.
The relative antenna-based delay correction between R and L was then derived using the primary calibrator.
The leakage terms between the R and L circularly polarized feeds were found from
the polarization leakage calibrator (Table~\ref{table:obs}).  Finally, 
the absolute polarization position angle
on the sky was determined from the known position angle for 
the primary calibrator ($\chi\,=\,33^\circ$ for 3C~286).  During each of these steps, all previously determined calibration
tables were applied on the fly.  Finally, all calibration tables,
including the self-calibration table determined from the corresponding total intensity map, were applied
to the RL and LR data.

\begin{figure*}[h]
   \centering
   \includegraphics*[width=3in]{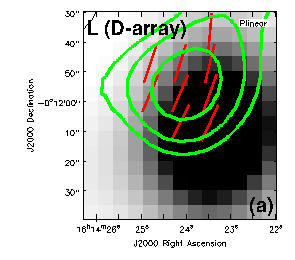}
   \hspace{0in}
   \includegraphics*[width=3in]{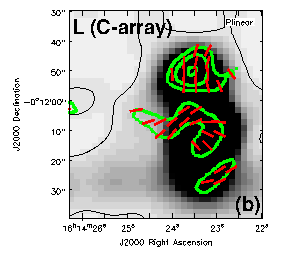}
   \hspace{0in}
   \includegraphics*[width=3in]{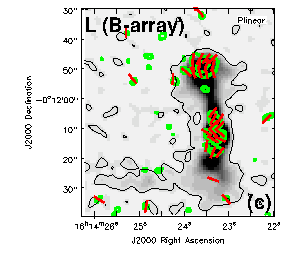}
   \caption{L-band polarization images of UGC~10288 (green contours) with E vectors rotated by 90$^\circ$ (red,
formed where Q and U $>$ 3$\sigma$).  
The greyscale represents the corresponding total intensity (I) image along with its
2$\sigma$ contour (black) from Fig.~\ref{fig:Lband}.  Map parameters
are given in Table~\ref{table:map-ugc10288}, except that the D-array polarization image has been
made from Q and U data that have been smoothed slightly to match the corresponding I data. 
 {\it (a)} D-array map; contours are at 105
(3$\sigma$), 150, and 220 $\mu$Jy beam$^{-1}$.
{\it (b)} C-array map; contours are at  65.1 (3$\sigma$), 110, and 175 $\mu$Jy beam$^{-1}$.
{\it (c)} B-array map; contours are at
38.1 (3$\sigma$), 90, and 175 $\mu$Jy beam$^{-1}$.}
\label{fig:Lband-pol}
\end{figure*}

\begin{figure*}[h]
   \centering
   \includegraphics*[width=3in]{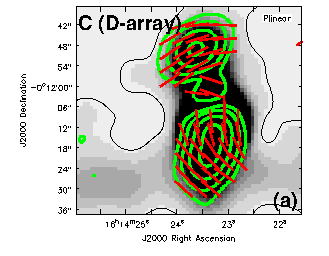}
   \hspace{0in}
   \includegraphics*[width=3in]{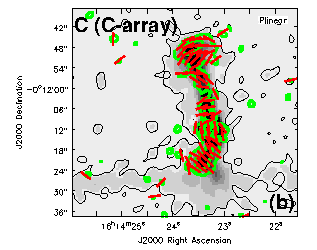}
   \caption{C-band polarization images of UGC~10288 (green contours) with E vectors rotated by 90$^\circ$ (red,
formed where Q and U $>$ 3$\sigma$).  
The greyscale represents the corresponding total intensity (I) image along with its
2$\sigma$ contour (black) from Fig.~\ref{fig:Cband}.  Map parameters
are given in Table~\ref{table:map-ugc10288}. 
 {\it (a)} D-array map; contours are at
16.2 (3$\sigma$), 35, 70, and 120 $\mu$Jy beam$^{-1}$.
{\it (b)} C-array map; contours are at
8.7 (3$\sigma$), 30, 75, and 140 $\mu$Jy beam$^{-1}$.
}
\label{fig:Cband-pol}
\end{figure*}

Imaging of Stokes Q and U proceeded in the same fashion as for I but the largest spatial scale was 
dropped during cleaning (except for D-array L-band) since the extent of the polarized emission is not as 
great as in total intensity.  Maps of polarized intensity were then formed for each array and frequency,
correcting for the bias introduced by the
fact that P images do not obey Gaussian statistics \citep{sim85,vai06}.
Maps of E vectors rotated by 90$^\circ$ were also formed after first blanking all emission less than
3$\sigma$, where $\sigma$ is the rms noise of the respective Q and U maps, resulting
in an uncertainty of $\pm$ 10 degrees. 
These vectors would be
equivalent to the intrinsic direction of the regular magnetic field, B, if there were
no Faraday rotation.  
The maps of polarized intensity (without primary beam corrections)
superimposed with the rotated E vectors are 
shown in
Figs.~\ref{fig:Lband-pol} and \ref{fig:Cband-pol}.

\subsection{Combined Array/Frequency Images and Spectral Index Map}
\label{sec:combined}

The uv data at each frequency {(B, C, and D arrays at L-band and C and D arrays at C-band)}
 were combined in order to obtain higher S/N and more complete uv coverage for
each of the bands.  The results of these combinations are parameterized in 
Table~\ref{table:combined} and each map was also corrected for the primary beam as described above.
We will refer to them as needed.

The power and versatility of CASA, moreover, along with the multi-array and frequency observations that we have
carried out also allow us to form a {\it single} image from {\it all} arrays and frequencies. The 
self-calibrated uv data 
from all five observations are input into the {\it clean} algorithm and, with a point-by-point fit of the spectral
index, 
an image can be made that corresponds to a frequency which is intermediate between L-band and C-band,
i.e. at $\nu\,=\,$4.13 GHz.  This process is time-consuming but
produces a final image with improved sensitivity to spatial scales and, on average, lower rms.  We show
the result in Fig.~\ref{fig:all_colour}, where the two images differ only in the uv weighting function adopted.

Since the spectral index, $\alpha$ ($I_\nu\,\propto\,\nu^\alpha$), is fitted during the imaging, a map of
$\alpha$ has also been formed (see Paper 2 for details)
 and is shown in Fig.~\ref{fig:alpha} along with its error map.  A 3$\sigma$ cutoff has been applied to
the data when forming the $\alpha$ map and the result has been corrected for the spectral index of the
primary beam. The error map represents random errors at each point as
described in \cite{rau11} (their Eqn. 39).
Note that similar spectral index maps were formed for
each individual data set as well (Sect.~\ref{sec:I_imaging}), 
but we show only the combined array/frequency result because of its
higher S/N.
In principle, it is also possible to solve for curvature in the spectral index (see Paper 2 for examples);
however, the uncertainties become large,
given the S/N in the disk of UGC~10288, and we did not pursue this for
UGC~10288. 

The remaining issue is to consider whether there might be missing flux from a lack of short
spacings. At L-band, we detect all spatial scales less than 16 arcmin and, at C-band, less than
4 arcmin. {The optical galaxy is 4.9 arcmin in diameter (Table~\ref{table:prop-ugc10288})
but, as illustrated in Figs.~\ref{fig:Lband} and \ref{fig:Cband}, the radio diameter does not extend
beyond $\approx$ 3 arcmin.  Moreover, we do not observe any `negative bowls' around the emission,
as would be the case if there were missing spacings.  Consequently, there should be no missing 
flux in these data.}

\section{Results}
\label{sec:results}

\subsection{The Total Intensity Emission}
A remarkable result of these observations is the discovery of a strong, double-lobed 
extra-galactic radio source (EGRS)
which is most certainly in the background of the UGC~10288 field.
Previous to these observations, the only radio continuum images obtained were of the galaxy and background
source blended together \citep[see][]{hum91,con98}.  
Indeed, the previously measured flux density from the combined sources,
$S_{1.4~GHz}$ = 26.1 mJy \citep{con98} met our minimum flux density criterion 
(23 mJy) for membership
in the CHANG-ES survey (Paper 1); however, it would have fallen far below that cut-off,
 had a measurement existed for the galaxy alone.

\begin{figure*}[h]
   \centering
   \includegraphics*[width=7in]{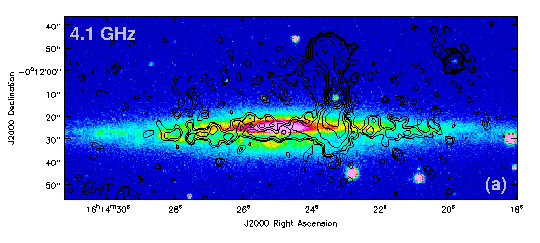}
   \hspace{-0.20in}
   \includegraphics*[width=7in]{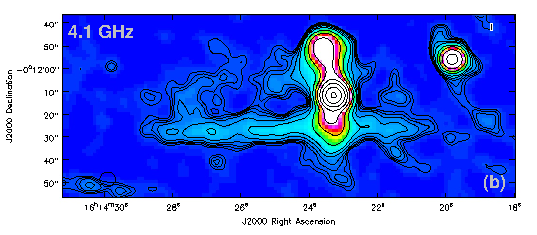}
   \caption{Total intensity combined array images of UGC~10288 (all arrays, both bands) 
corresponding to a frequency, 
$\nu_0\,=\,4.1$ GHz, shown at two different uv weightings.  
{\it (a)}  I contours of the higher resolution Briggs robust = 0 uv weighting at 
14 (2$\sigma$), 20, 30, 56, 100, 200, 400, 800, 2000, and 4000 $\mu$Jy beam$^{-1}$
over the Sloan Digital Sky Survey (SDSS) r band colour image. The beam size
($\sim$ 3.5 arcsec, Table~\ref{table:combined}) is shown at lower left.
{\it (b)} I contours and colour of the lower resolution Briggs robust = 0 + 16 k$\lambda$ uv tapered image, with
contours at 20 (2$\sigma$), 25, 30, 38, 55, 80, 160, 300, 600, 1200, 2400, 4800, and 8500 $\mu$Jy beam$^{-1}$. The beam size ($\sim$ 6.5 arcsec, Table~\ref{table:combined}) is shown at lower left.
}
\label{fig:all_colour}
\end{figure*}

We now see both the galaxy and the background source in detail.  There is an optical point
source at the core of the EGRS (Fig.~\ref{fig:all_colour}), with
Sloan Digital Sky Survey (SDSS) DR9 
identifer, J161423.28-001211.8, listed as a galaxy with a photometric redshift of
$z_{phot}$ = 0.388 $\pm$ 0.026.
We have named this background source, {\it CHANG-ES A},
and labelled it in Fig.~\ref{fig:Lband}b. 
 Another strong source ({\it CHANG-ES B}, labelled in Fig.~\ref{fig:spitzer}) is blended
with the disk of UGC~10288 in D-array L-band (Fig.~\ref{fig:Lband}a), making the UGC~10288
disk appear to
be skewed towards the north-west in that figure.  {\it CHANG-ES B} also has an optical
background galaxy at its center, namely J161419.79-001155.6 at $z_{phot}$ = 0.39 $\pm$ 0.11,
 in agreement with the redshift of {\it CHANG-ES A}.  Therefore {\it CHANG-ES A} and
{\it CHANG-ES B} possibly form a background pair.

\subsubsection{Flux Densities and Global Spectral Index}
\label{sec:flux}

Estimates of the flux densities of the various sources are given in Table~\ref{table:derived-ugc10288}.
All flux densities were measured from the primary-beam corrected 
C-array L-band (Fig.~\ref{fig:Lband}b)
and D-array C-band (Fig.~\ref{fig:Cband}a)
maps.  For both bands, the total flux density of
UGC~10288 and {\it CHANG-ES A} were measured together.  Then the flux density of {\it CHANG-ES A}
was estimated from a measurement within the two dashed lines 
shown in Fig.~\ref{fig:Cband}a, a width of 34 arcsec.  The flux density
of UGC~10288 alone was then measured in a region of equivalent width
to the east of {\it CHANG-ES A} along the galaxy's disk.  Assuming that this measured flux
density
is approximately equivalent to the flux density of UGC~10288 
{\it within} the box occupied by {\it CHANG-ES A}, it
 was then
subtracted from the {\it CHANG-ES A} measurement as a correction for the contribution of
the galaxy's disk within the dashed lines. At L-band, this correction amounted to
4\% of the flux density of {\it CHANG-ES A}, and 22\% of the flux density of UGC~10288.
At C-band, the correction constituted 4\% of the flux density of {\it CHANG-ES A} and 
32\% of the flux density of UGC~10288. 
In addition, at L-band,
the totals were then summed and compared to the combined, blended emission from
all three sources, i.e.
UGC~10288, {\it CHANG-ES A} and {\it CHANG-ES B} 
as measured in the D-array L-band image (Fig.~\ref{fig:Lband}a);
the totals agree to within 0.5\%.

The global spectral index of UGC~10288 was calculated from the flux densities as
just described, the result given in Table~\ref{table:derived-ugc10288}.  The two
sources are both clearly dominated by non-thermal emission.

\subsubsection{The Star Formation Rate of UGC~10288}
\label{sec:sfr}

It is remarkable that UGC~10288 reveals extra-planar emission 
(see next two subsections) when the total
flux density is only a few mJy. This leads us to consider the star
formation rate (SFR) of the galaxy, since halo emission is often associated
with such activity.

UGC~10288's SFR$_{IRAS}$ of 1.3 M$_\odot$ {yr$^{-1}$}  as given in
Table~\ref{table:prop-ugc10288}, has been calculated from Infrared Astronomical
Satellite (IRAS)
data (Paper 1), but, like the radio continuum, the IR emission will 
certainly be
contaminated by emission from {\it CHANG-ES A}. 

A closer inspection of the IRAS flux densities, moreover, suggests
foreground contamination as well.
Given the observed IRAS 60 to 100 $\mu$m flux density ratio of UGC~10288
($f_{(60\,\mu m)}/f_{(100\,\mu m)}$ = 0.236), the galaxy appears to be
extremely cold. 
In fact, this flux density ratio is $\approx$20\% smaller (colder) than that for
the most extreme (coldest) galaxy included in the Dale \& Helou (2002) spectral
energy distribution (SED) models of normal star-forming galaxies. 
If {\it CHANG-ES A} is indeed a background radio galaxy, we naively would not expect it to
be colder than all galaxies in the local Universe. 
If anything, we would expect a warmer temperature arising from hot dust being
powered by the accreting black hole. 
Therefore, it seems that the most likely explanation for the cold dust
temperature is significant contamination by difficult-to-remove foreground cirrus
emission that is contributing to the IRAS flux densities. 

Given these issues,
an alternative approach is to use the calibrated, known radio continuum - FIR
relation to estimate the SFR of UGC~10288 from its
 radio continuum emission as estimated in Sect.~\ref{sec:flux} with {\it CHANG-ES A}'s
flux density subtracted.
 For this, we adopt the
calibrations of \cite{mur11}
which use a Kroupa initial mass function (IMF) and a supernova cutoff of 8
M$_\odot$.  The SFR estimated from
the thermal radio continuum requires an electron temperature which we take to be
10$^4$ K. The SFR
estimated from the non-thermal radio continuum requires a knowledge of the
non-thermal spectral index,
$\alpha_{NT}$, whereas we have, from the observed flux densities of
Table~\ref{table:derived-ugc10288}, a global spectral index
of $\alpha\,=\,-0.76$, using the convention,
 ($S_\nu\,\propto\,\nu^\alpha$)\footnote{This is opposite to that of
\cite{mur11}.}.

However, we can estimate $\alpha_{NT}$ by combining the thermal and non-thermal
calibrations
of \cite{mur11} (their Eqns. 11 and 14) and the constraint that the measured
flux density is the sum of the non-thermal
and thermal flux densities at any frequency. i.e.
\begin{equation}
\frac{S_{\nu_1}}{S_{\nu_2}}\,=\,\left(\frac{\nu_1}{\nu_2}\right)^{\alpha_{NT}}
\frac{
\left[0.144\,(\nu_1)^{-\alpha_{NT}-0.1}\,+\,1\
\right]}
{\left[0.144\,(\nu_2)^{-\alpha_{NT}-0.1}\,+\,1\
\right]}
\label{eqn:alpha-solve}
\end{equation} 
where $S_\nu$ is the total flux density at frequency, $\nu$ and these calibrations assume
a linear dependence of SFR on both thermal and non-thermal luminosity.  The use of this
equation infers a thermal/non-thermal fraction which has a dependence on both frequency and
$\alpha_{NT}$.

Using our measured flux densities of Table~\ref{table:derived-ugc10288} at two
frequencies, we can solve
Eqn.~\ref{eqn:alpha-solve} to find $\alpha_{NT}\,=\,-1.04$. 
This leads to a thermal to total and non-thermal to total flux density in L-band
of,
${(S_{T}/S)}_{1.5}\,=\,0.17$, ${(S_{NT}/S)}_{1.5}\,=\,0.83$, respectively.  In
C-band, the results are
${(S_{T}/S)}_{6.0}\,=\,0.44$, ${(S_{NT}/S)}_{6.0}\,=\,0.56$, respectively.

Finally, with the SFR calibration \citep[eqn. 15 of][]{mur11}, we find SFR =
0.51 M$_\odot$ yr$^{-1}$.

We caution that, if the non-thermal luminosity,
$L_{NT}\,\propto\,$SFR$^n$, where $n\,>\,1$ \citep[e.g. n$\,=\,1.3$][]{nik97},
then there would be a weak dependence on SFR on the right hand side of
Eqn.~\ref{eqn:alpha-solve}; the adjustment, however, does not significantly 
perturb the above results\footnote{If, for example, the constant of proportionality
for the Niklas \& Beck non-thermal relation is the same as for the linear one used here, then 
one would replace both `ones' on the right hand side 
of Eqn.~\ref{eqn:alpha-solve} with SFR$^{0.3}$. This would result in $\alpha_{NT}\,=\,-1.14$,
or a change of
9\% using SFR = 0.5 M$_\odot$ yr$^{-1}$.  The SFR, in turn, would change by 3\%.}.

This newly determined SFR is considerably lower than the original SFR which was found from IRAS
values and should be
 more accurate since it corrects for the contribution of {\it CHANG-ES A} and also uses a 
spectral index determined from the data.
If, instead, we use the global FIR-radio continuum relation \citep[Eqn. 17
of][]{mur11} based on earlier
calibrations \citep{dej85,hel85}, then
SFR = 0.39 M$_\odot$ yr$^{-1}$. 
In either case, the radio-derived SFRs are clearly much lower than the
IRAS-based estimate {of 1.3 M$_\odot$ yr$^{-1}$}.

In summary, the SFR based on IRAS photometry (Table~\ref{table:prop-ugc10288}) 
neither corrects for the {strong}
contribution from {\it CHANG-ES A} nor for contamination from foreground cirrus.  We
estimate that a value between 0.4 and 0.5 M$_\odot$ yr$^{-1}$ is {a more accurate estimate
of the true SFR in UGC~10288.  Note that the range given here is dominated by the assumptions
that have been made, rather than by a formal error determination.}
This galaxy is now one of the lower SFR galaxies in the CHANG-ES
sample (Paper 1).


\subsubsection{Foreground/Background Point-like Sources}
\label{sec:pointsources}

\begin{figure*}[h]
   \centering
   \includegraphics*[width=7in]{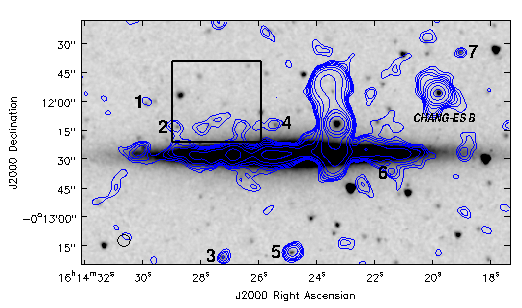}
   \caption{
Combined C and D array C-band contours over a greyscale $\lambda\,$3.6 $\mu$m
map from the Spitzer telescope.  Contour levels are at
8 (2$\sigma$), 11, 15, 20, 30, 50, 100, 300, 600, 1200, and 3000 $\mu$Jy beam$^{-1}$, and the
beam size ($\sim$ 6 arcsec, Table~\ref{table:combined}) 
is shown at lower left.  Several point sources showing radio emission that
are distinct from the disk of UGC~10288 have been
numbered (Table~\ref{table:pointsources}). The box encloses a region containing the radio continuum
extension seen in Fig.~\ref{fig:all_colour}b.
}
\label{fig:spitzer}
\end{figure*}

Fig.~\ref{fig:spitzer} shows the combined C+D array C-band map of UGC~10288
overlaid on the $\lambda$ 3.6 $\mu$m image from the Spitzer Space 
Telescope\footnote{The `post-Basic Calibrated Data' (post-BCD) were used.}.
These radio data have the lowest rms of all
combined data sets (Table~\ref{table:combined}) as well as high resolution.
Comparing with the Spitzer image, which was chosen 
because it should show both foreground stars as well as background
quasars, should reveal how much radio emission (especially in regions of
apparent halo features) is associated with
point-like sources in the displayed field.

We searched for sources for which both the IR and radio emission are point-like
and for which the
radio emission is sufficiently
separated from the disk of UGC~10288 to allow a discrete measurement of flux density.  The
process was then repeated for the
combined B+C array L-band map (not shown), although
the latter image has almost 4 times higher rms noise than at C-band for
approximately equal beam sizes.
The resulting sources are labelled in 
Fig.~\ref{fig:spitzer} and their properties listed, along with SDSS identifiers (where
available) in Table~\ref{table:pointsources}. 

Let us consider the region that is encompassed by the box shown in
Fig.~\ref{fig:spitzer} 
(16$^{\rm h}$ 14$^{\rm m}$ 25.9$^{\rm s}$ $\le$ RA $\le$ 16$^{\rm h}$ 14$^{\rm m}$ 28.9$^{\rm s}$;
-00$^\circ$ 12$^\prime$ 21$^{\prime\prime}$ $\le$ DEC $\le$ -00$^\circ$ 11$^\prime$ 40$^{\prime\prime}$). 
This box contains significant `halo' emission in the form of a large extended
arc-like feature visible 
 in  the all-data image of
Fig.~\ref{fig:all_colour}b.  
This arc extends to a height of 
21 arcsec (3.5 kpc) above the plane.
The flux density contained within this box is 
$S_{4.1~{\rm GHz}}\,=\,450~\pm~80$ $\mu$Jy\footnote{Measured from the primary beam corrected
image. The error is measured as $\Delta\,S\,=\,rms * \sqrt{N_B}$, where $N_B$ is the number
of independent beams in the region.}. 
Only one identified point-like source (Source 2) is located within this region whose
flux density, extrapolated to 4.1 GHz, is $<$ 16 $\mu$Jy (Table~\ref{table:pointsources}).
Consequently, although we cannot conclude that the arc-like feature represents a single
coherent structure, its flux density does not result from background sources.
Hereafter, we will refer to this features as the `arc' for simplicity.

There is also a feature extending to the north on the west side of the galaxy
(at RA $\approx$ 16$^{\rm h}$ 14$^{\rm m}$ 21.5$^{\rm s}$, DEC $\approx$ -00$^\circ$ 12$^\prime$ 12$^{\prime\prime}$) 
as shown in
Fig.~\ref{fig:all_colour}b; but there are no point sources meeting our criteria
in this region at all.

We caution that if a background source displays double-lobed radio structure,
then it could be displaced with respect to the IR source and
might have been missed in this search, as would objects
with weak IR flux and those that are blended with the disk.  However, even if we increased our
point-like source flux density estimate by a factor of two, its contribution 
 to the flux density of halo features in UGC~10288 would still be negligible.
By including other array configurations in the maps, in particular short spacings, the
sensitivity to extended halo features is enhanced but there should be no change in
sensitivity to point sources.  

It is beyond the scope of this paper to examine in detail the nature of the point-like sources.
SDSS identifiers, where available, indicate that three of the four identified sources
are stars which would be an interesting result given the detected radio emission. 
However, one of these sources (Source 5,
SDSS~J161424.84-001318.3) although labelled as a star, 
has SDSS colours (u-g = +0.18, g-r = -0.18, r-i = 0.45,
i-z = -0.07) which are entirely consistent with those of a quasar over a range of
 redshift \citep[][their Fig. 28]{par12}. 

\subsubsection{Comparison with Images at other Wavebands}
\label{sec:overlays}

It is apparent that halo radio emission is
associated with UGC~10288.  Here, we
appeal to
ancillary data, where available, to 
 shed further light on the high latitude features.

\begin{figure*}[h]
   \centering
   \includegraphics*[width=7in]{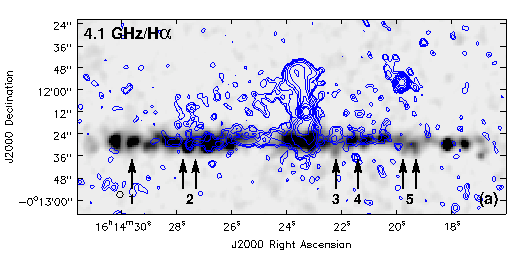}
   \hspace{-0.40in}
   \includegraphics*[width=7in]{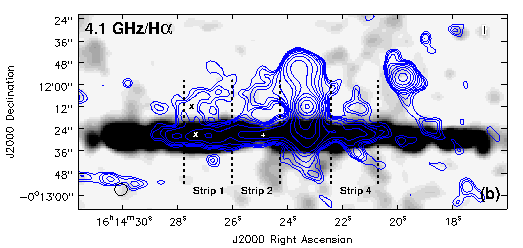}
   \caption{Contours from Fig.~\ref{fig:all_colour} over the H$\alpha$ emission measure map of
\cite{ran96} smoothed to the same resolution as the radio data
in each case.  The beam size is shown at bottom left.
{\it (a)} Contours are from Fig.~\ref{fig:all_colour}a and the resolution is
$\sim$ 3.5 arcsec (Table~\ref{table:combined}).  The arrows point to features discussed in
Sect.~\ref{sec:overlays}.{\it (b)} Contours are from Fig.~\ref{fig:all_colour}b and the
resolution is $\sim$ 6.5 arcsec, Table~\ref{table:combined}).  A plus marks the center of
UGC~10288 and two crosses mark the positions discussed in Sect.~\ref{sec:pol}.
The labelled strips are discussed in Sect.~\ref{sec:overlays}.
}
\label{fig:overlays_a}
\end{figure*}

\begin{figure*}[h]
   \centering
   \includegraphics*[width=7in]{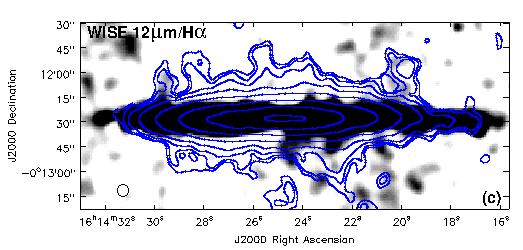}
  \hspace{-0.40in}
   \includegraphics*[width=7in]{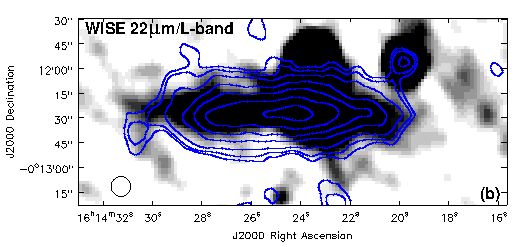}
   \caption{{\it (a)} Contours (in arbitrary units) of the WISE $\lambda\,$12
$\mu$m image over the H$\alpha$ image of
\cite{ran96}, both smoothed to 7.0 arcsec resolution (beam shown at lower left). 
 {\it (b)} Contours (in arbitrary units) of the WISE $\lambda\,$22
$\mu$m image over the combined BCD array L-band data (original information given in
Table~\ref{table:combined}) smoothed to the same 12 arcsec
resolution as the WISE data (beam at lower left).
}
\label{fig:overlays_b}
\end{figure*}

Few resolved images of UGC~10288 are available at other wavebands but, of those
available, we show several overlays in
Figs.~\ref{fig:overlays_a} and \ref{fig:overlays_b}.  Fig.~\ref{fig:overlays_a} shows
our all-array and all-frequency radio continuum images
of Fig.~\ref{fig:all_colour} superimposed on
the H$\alpha$ image of \cite{ran96}.  In Fig.~\ref{fig:overlays_b}, we show contours
from the Wide-field Infrared Survey Explorer (WISE) \citep{wri10}
at $\lambda\,$12 $\mu$m
and $\lambda\,$22 $\mu$m over H$\alpha$ emission and our L-band combined array image,
respectively.  In each overlay, the spatial resolutions of the two images are identical. 

Because radio emission from UGC~10288 is blended with that of {\it CHANG-ES A}, it is difficult
to disentangle which source is responsible for specific features observed at locations close
(in projection)
to that background source.  For example, a horizontal feature appears to emanate from the
northern lobe of {\it CHANG-ES A} towards the east 
(see RA $\sim$ 16$^{\rm h}$ 14${\rm m}$ 25$^{\rm s}$, DEC ~$\sim$ -00$^\circ$ 12$^\prime$ 06$^{\prime\prime}$
in Fig.~\ref{fig:overlays_a}b).
There is also a projection which extends to the south of {\it CHANG-ES A}'s southern radio lobe
(RA $\sim$ 16$^{\rm h}$ 14${\rm m}$ 235$^{\rm s}$, DEC ~$\sim$ -00$^\circ$ 12$^\prime$ 43$^{\prime\prime}$
in Fig.~\ref{fig:overlays_a}b) and farther south as a narrower radio extension
in Fig.~\ref{fig:overlays_b}b).
 Neither of these have obvious counterparts at other wavebands,
although the southern feature appears to lie between two H$\alpha$ extensions (see below).

The IR emission does not distinguish between UGC~10288 and background sources either, since
 for example, the bright background radio source, {\it CHANG-ES B}, to the west of
{\it CHANGES-A} (marked in Fig.~\ref{fig:spitzer}) is also an IR emitter at both 
$\lambda\,$12 and 22 $\mu$m. 

As we have no way to determine the origin of radio continuum features near {\it CHANG-ES A}, we will
concentrate instead on other features that are farther from this region and 
are more likely to be associated with
UGC~10288.  The
 H$\alpha$ emission is helpful in this regard, since it clearly has an origin in UGC~10288 itself. 

We must also note that each of the images suffers from some 
low-level artifacts.  

For the radio 
images, the map contains residual sidelobes which,
for the most part, 
we have included in measuring and quoting rms map values.
However, residual sidelobes from a source 
6.3 arcmin  
 to the east of UGC~10288
are likely responsible
for the horizontal feature at the south-east side of Fig.~\ref{fig:overlays_a}b
(RA $\sim$ 16$^{\rm h}$ 14$^{\rm m}$ 30$^{\rm s}$, DEC ~$\sim$ -00$^\circ$ 12$^\prime$ 50$^{\prime\prime}$),
and residual sidelobes from another point source 6.2 arcmin to the south-west
may be responsible for the faint greyscale NE-SW feature visible on the west side of
Fig.~\ref{fig:overlays_b}b 
(RA $\sim$ 16$^{\rm h}$ 14$^{\rm m}$ 17$^{\rm s}$, DEC ~$\sim$ -00$^\circ$ 12$^\prime$ 15$^{\prime\prime}$). 

Similarly, 
in the H$\alpha$ image, it has been noted that the galaxy's
bulge has been oversubtracted \citep{ran96}. And both WISE images show broad
(arcmin scale) NE-SW striping which may be producing or contributing to the broad
southward extension at
RA $\sim$ 16$^{\rm h}$ 14$^{\rm m}$ 30$^{\rm s}$, 
DEC ~$\sim$ -00$^\circ$ 12$^\prime$ 50$^{\prime\prime}$ (Fig.~\ref{fig:overlays_b}a).

In any following discussion, therefore, we will keep these cautions in mind
and discuss results that appear to be robust to such artifacts.
Fig.~\ref{fig:overlays_a}a shows the highest resolution ($\sim$ 3.5 arcsec) 
radio/H$\alpha$ overlay with arrows pointing to specific features.  
Discrete 
in-disk H$\alpha$ 
emission reveals specific sites of unobscured star formation.

Feature 1 is a discrete star forming region in the disk above which there is a vertical plume
of H$\alpha$ emission (to the north).  The H$\alpha$ plume is more readily
seen in the smoothed map of Fig.~\ref{fig:overlays_a}b and has a 
$\lambda\,$12 $\mu$m counterpart (Fig.~\ref{fig:overlays_b}a). No radio emission
can be confirmed in this plume, but a slight extension visible in
L-band greyscale image of Fig.~\ref{fig:overlays_b}b is close to its location, slightly
displaced to the east\footnote{This feature is just below point source \#1
(Fig.~\ref{fig:spitzer} and Table~\ref{table:pointsources}).}.  
That is, the radio extension's location is
more closely aligned with the gap\footnote{By `gap', we do not mean the absence
of H$\alpha$ emission but rather a sharp drop in the emission compared to adjacent
locations.} immediately
 to the east of star forming region \#1.

Feature 2 refers to several star forming complexes that are located below the 
strong northward radio plume seen in Fig.~\ref{fig:overlays_a}.  The two arrows point
to gaps in the H$\alpha$ in-disk emission, the easternmost one also corresponding
to a gap in the in-disk radio emission.  Two `prongs' in the radio emission
close to the disk (Fig.~\ref{fig:overlays_a}b) are immediately above these gaps.

The southern H$\alpha$ extension of feature 3 was originally pointed out 
by \cite{ran96} and is in a region of low brightness in comparison to the star forming
regions
to the east and west of it.
This feature shows a small northwards elongation in the radio
at high resolution (Fig.~\ref{fig:overlays_a}a).

Feature 4 points to another gap in the H$\alpha$ emission between two star forming regions.
There is a small radio extension to the north that is visible in the high
resolution image (Fig.~\ref{fig:overlays_a}a) but a much larger
corresponding radio continuum feature can be
seen at lower resolution  (Fig.~\ref{fig:overlays_a}b).

Finally, feature 5 refers to two southwards prongs which were originally
identified by \cite{ran96} who also noted a `hemispherical' shape formed
by these two features together.  We have not detected any high-latitude radio 
emission associated with this feature. 

In summary, our high resolution data suggest that radio continuum features are more likely
to be seen above in-disk gaps in the H$\alpha$ emission.

It should be pointed out, of course, that apparent correlations above an edge-on
disk could apply to any location along the line of sight (although such
indeterminacy is less severe at large galactocentric radii). 
This means that such apparent correlations do not necessarily represent physical correlations.
Nevertheless, the observed trend is worth pointing out, since it is rare that an edge-on galaxy
has adequate data 
to compare H$\alpha$ and radio continuum results at the same resolution with good sensitivity.
A similar anti-correlation has been observed before  
 between in-disk H$\alpha$ emission and high
latitude radio continuum features in NGC~5775 \citep{lee01}.
We discuss this further in
Sect.~\ref{sec:discussion}.

On broader scales, it is more difficult to associate high latitude radio emission
with in-disk activity.
 Intriguing features are a south-wards H$\alpha$ arc seen in
Fig.~\ref{fig:overlays_a}b at RA $\sim$ 16$^{\rm h}$ 14$^{\rm m}$ 24.5$^{\rm s}$,
DEC $\sim$ -00$^\circ$ 12$^\prime$ 48$^{\prime\prime}$, and a possible
companion feature at RA $\sim$ 16$^{\rm h}$ 14$^{\rm m}$ 22.7$^{\rm s}$, 
DEC $\sim$ -00$^\circ$ 12$^\prime$ 42$^{\prime\prime}$ between
which the southern radio continuum extension, mentioned above, lies.
In Fig.~\ref{fig:overlays_b}a, the $\lambda\,$12 $\mu$m emission shows
several other
 northwards extensions as well as a south-wards
 arc-like feature
on the east side of the galaxy, and in
Fig.~\ref{fig:overlays_b}b,
 and a southwards
radio extension seen at L-band on the far east end of the disk.  

We note two other observations regarding the distributions.

The first is the north-south radio asymmetry observed on the east side of
the major axis, best seen in Fig.~\ref{fig:overlays_a}b.
 i.e. there is a large arc-like extension towards the
north but nothing equivalent to the south\footnote{This is {\it not}
a result of sidelobes or negative bowls in the emission.}.  This
may reflect N/S asymmetries in star forming regions and/or 
N/S density asymmetries in the disk.  Such asymmetries can also be
produced by ram pressure as a galaxy moves through an 
 intergalactic medium;  this is possible for UGC~10288, given that it is in a sparse
group (Sect.~\ref{sec:introduction}), but we have no independent 
observations that would argue for or against this speculation.

The second {observation} is that UGC~10288 does not appear to have a global radio halo, but rather
it has discrete high latitude radio extensions. 
Table~\ref{table:scaleheights} compares beam-corrected
exponential scale heights between 
the 4.1 GHz all-data map (Fig.~\ref{fig:all_colour}b) and both the H$\alpha$ and
WISE $\lambda\,$12 $\mu$m data at the same (7 arcsec) resolution
according to the method of \cite{dum95}.  The fitting has been carried out for data that
have been averaged in strips,
each  26.4" = 1.76 s wide, that are labelled in Fig.~\ref{fig:overlays_a}b\footnote{Strip \#3
has been avoided since it contains {\it CHANG-ES A}.}.  
The values
are from single-exponential fits that apply to
the disk (for example, for Strip \#2 North, the horizontal feature at DEC $\sim$ -00 12 06, was
avoided).  Note that the strips that do not have significant extensions
(mainly \#2 South but also \#1 South and \#2 North) 
have low to modest scale heights in radio continuum
(0.46, 0.85 \& 0.86 kpc, respectively), whereas the strips containing
discrete extensions (\#1 North \& \# 4 North/South) are wider in z, on average
(1.3, 1.4, \& 1.1 kpc, respectively) as would be expected.  See Sect.~\ref{sec:discussion} for
more discussion.



\begin{figure*}[h]
   \centering
   \includegraphics*[width=7in]{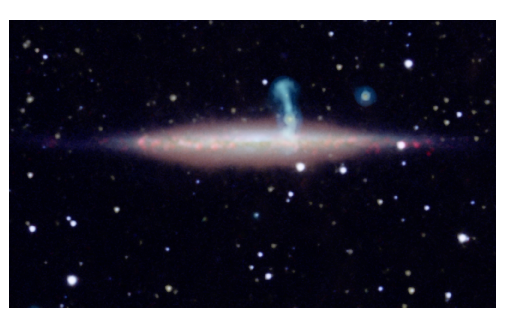}
   \caption{Multi-frequency
colour image of the UGC~10288 field showing our C-array C-band image
(Fig.~\ref{fig:Cband}b) in cyan, the combined all-array, all-frequency image
in darker cyan (Fig.~\ref{fig:all_colour}b),
the WISE $\lambda\,12~\mu$m image
(Fig.~\ref{fig:overlays_b}a) in orange, the Spitzer $\lambda\,3.6~\mu$m image
(Fig.~\ref{fig:spitzer}) in yellow, the H$\alpha$ image of \cite{ran96} 
Fig.~\ref{fig:overlays_a} in rose, the SDSS r band image
(Fig.~\ref{fig:all_colour}a) in blue and the SDSS g band image in
purple.  Smoothing, colour choice and masking follow the techniques described in \cite{rec07}.
Spatial resolutions vary and have been chosen for visual effect.
 Both {\it CHANG-ES A} and {\it CHANG-ES B} (cyan and darker cyan at lower resolution) 
are quite obvious, and the large vertical width of the disk which appears salmon coloured at the edges,
is due to the large width seen in the  WISE $\lambda\,12~\mu$m image.
}
\label{fig:jayanne}
\end{figure*}

In order to give an overview of the relative spatial distributions of the various
sources of emission,
we finally display a multi-frequency image (Fig.~\ref{fig:jayanne}) in which several
data sets have been overlayed with different colours (see caption). Our radio data show
{\it CHANG-ES A} and {\it CHANG-ES B} in cyan; our masking technique \citep{rec07}
allows us to show their optical counterparts (yellowish) at their cores.
The disk of UGC~10288 looks quite wide vertically in this picture because of the
WISE $\lambda\,12~\mu$m image.  
For more information as to the techniques used to
combine the data sets, see \cite{rec07}.

\subsection{Spectral Index}
\label{results:spectral_index}

\begin{figure*}[h]
   \centering
   \includegraphics*[width=6.5in]{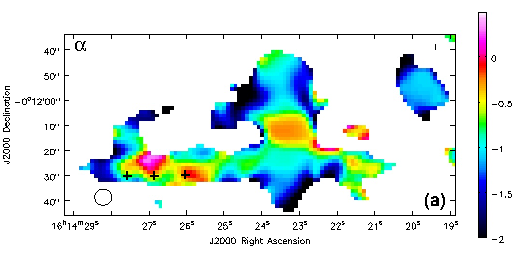}
   \hspace{-0.20in}
   \includegraphics*[width=6.5in]{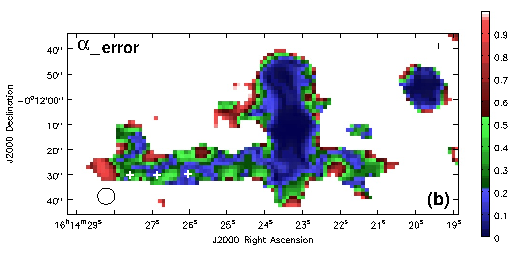}
   \caption{Spectral index map, $\alpha$ {\it (a)}, and uncertainty in the
spectral index {\it (b)} (random errors only), 
 made from all data (all arrays, both bands) corresponding to
the uv weighting image of Fig.~\ref{fig:all_colour}b ($S_\nu\,\propto\,\nu^{\alpha}$) with a 3$\sigma$
intensity cutoff applied.  The 3 brightest star forming regions from Fig.~\ref{fig:overlays_a}a
have been marked with crosses and the beam is shown at lower left.
 Several points
from the $\alpha$ map have uncertainties that appear white because they have values that are $>$ 1. 
The beam size ($\sim$ 6.5 arcsec, Table~\ref{table:combined}) is shown at lower left.
}
\label{fig:alpha}
\end{figure*}

The spectral index map along with a point-by-point error map (see Sect.~\ref{sec:combined}) are shown
in Fig.~\ref{fig:alpha}.  The disk of UGC~10288 is of course blended with {\it CHANG-ES A} which shows
the classic signature of a double-lobed radio source:  a flatter spectral index at the center
and steeper in the lobes.

The spectral index in the disk of UGC~10288 shows complex structure but, as also
shown by the global spectral index (Sect.~\ref{sec:flux}), non-thermal spectra
dominate (Table~\ref{table:derived-ugc10288}).  
At the locations of the three brightest star forming regions in the disk (denoted with crosses),
 as identified from the
H$\alpha$ map, the spectral indices are 
$\alpha\,=\,-0.7~\pm~0.3$,      
$\alpha\,=\,-0.4~\pm~0.2$, and  
$\alpha\,=\,-0.1~\pm~0.3$       
from east to west, respectively.  If averaged over a beam rather than measured at the H$\alpha$ peak, these values become
$\alpha\,=\,-0.8~\pm~0.3$, 
$\alpha\,=\,-0.4~\pm~0.2$, 
and
$\alpha\,=\,-0.3~\pm~0.2$ 
from east to west, respectively. Thus, only the westernmost
SF region is consistent with a purely thermal spectrum at its peak.  However, if we insist that the SFR 
calibrations discussed in Sect.~\ref{sec:sfr} hold at every point, then a flatter
observed spectral index actually implies a flatter non-thermal spectral index,
presumably 
reflecting a younger cosmic ray electron population.  Such details will be explored in future
CHANG-ES papers.

Note that these 3 star forming
regions do not align exactly with the flattest spectral indices.  The region with the flattest
spectral index is directly north of the central star forming region
 and there is another region of flatter spectral
 to the south-west of the
westernmost star forming region.  
This could be a result of outflows in these regions; such spectral index flattenings have been
observed before in regions of outflows \citep[e.g. NGC~5775,][]{lee01}.  
However, it is important to keep in mind
the uncertainties on these quantities.  Generally, the errors also increase towards the edges of the 
map.

\subsection{Polarization}

\subsubsection{Polarization Images and Sample Magnetic field}
\label{sec:pol}

Figs.~\ref{fig:Lband-pol} and \ref{fig:Cband-pol} 
show the polarization maps at L-band and C-band, respectively.
We do not detect polarization in the disk of UGC~10288 in the individual 
observations.  Rather,
it is the background radio source, {\it CHANG-ES A}, that shows polarization.  
Although a more complete analysis of the polarization-related properties
awaits a future paper, a few results are presented here.

For both 
 B-array, L-band (resolution $\sim$ 3.6 arcsec) and C-array, C-band
 (resolution $\sim$ 3.0 arcsec) the highest polarization
occurs right at the core of {\it CHANG-ES A}
(peak values and resolutions provided in Table~\ref{table:map-ugc10288}).
At
this location, the percentage polarization at L and C bands is 1.9\%
and 2.0\%, respectively.

North and south of the core, although the polarized intensity is lower, the
percentage polarization increases.  
At the peak polarization position of the C-band data in the
northern lobe, for example,
the values are
17\% and 30\% at L and C bands, respectively
(RA = 16$^{\rm h}$ 14$^{\rm m}$ 23.7$^{\rm s}$, 
DEC = -00$^\circ$ 11$^\prime$ 48$^{\prime\prime}$).  At the southern peak position, 
the corresponding values are 6.3\% and 31\% at L and C bands,
respectively (RA = 16$^{\rm h}$ 14$^{\rm m}$ 23.47$^{\rm s}$, 
DEC = -00$^\circ$ 12$^\prime$ 21$^{\prime\prime}$)\footnote{This southern peak position is not as far from
the core as is the northern peak position noted above. 
Since the southern emission is blended with the disk of UGC~10288, it is not clear whether the
southern peak position is in the jet or the lobe.}.

The L-band percentage polarization is weaker than at C-band, as expected, since more
Faraday de-polarization occurs at the lower frequency,
either intrinsic to the source or contributed via UGC~10288 or both.  {Moreover, the L-band
value is lower still for the southern lobe, compared to the northern, likely because of stronger
Faraday depolarization in the (possibly turbulent) disk of UGC~10288.}
The fractional polarization in the core is much lower than in either lobe; this
is not unusual for double-lobed radio sources
\citep[e.g.][]{rud86,sai98}
suggesting that much of the core/lobe difference is intrinsic to {\it CHANG-ES A}.

UGC~10288 does not lend itself well to global magnetic field calculations, given
the interfering presence of {\it CHANG-ES A}.  However, we have taken two sample positions, one in
the disk, and one in the  large northern arc on the east side of UGC~10288 visible in
the 4.1 GHz map,
 to estimate the 
total minimum energy magnetic field strength, $B$.
We use the revised minimum energy formulae given in \cite{bec05} for isotropic fields and adopt a
proton-to-electron number density ratio of 100.
The positions are marked with
crosses in Fig.~\ref{fig:overlays_a}b, the disk position being immediately below the arc.





At the position in the eastern disk,
a projected distance of $x\,=\,$39 arcsec
from the center,
 $\alpha\,=\,-0.44\,\pm\,0.18$ and
$I_{4.1~GHz}\,=\,70.4$ $\mu$Jy beam$^{-1}$.
Then
$\alpha_{NT}\,=\,-0.54$ 
and $I_{NT}\,=\,55.4$ $\mu$Jy beam$^{-1}$ by Eqn.~\ref{eqn:alpha-solve} and the arguments presented in
Sect.~\ref{sec:sfr}.  
For a line of sight distance of 6.0 kpc (estimated from a 1/e radial scale length) then the
minimum energy magnetic field strength is $B\,=\,10\,\pm\, 3$ $\mu$G.  {Here, the error represents
measurement error only, which is dominated by the uncertainty in $\alpha$; it does not include
errors that may be associated with the assumptions.}



 
At the position in the northern extension,
15 arcsec above the plane (2.5 kpc in projection),
$\alpha\,=\,-1.00\,\pm\,0.50$ (Fig.~\ref{fig:alpha}),
$I_{4.1~GHz}\,=\,36.0$ $\mu$Jy beam$^{-1}$, we assume that the emission is
dominantly non-thermal
based on the absence of H$\alpha$ emission, and take $l\,=\,1.5$ kpc (9 arcsec) which is
approximately the width of the arc at this location.  Then, provided that the conditions for
minimum energy are met for this region of steep spectral index, 
$B\,=\,12\,\pm\,5$   
$\mu$Gauss, {where, again, the error represents measurement error.}

As Fig.~\ref{fig:alpha} illustrates, there is considerable variation in spectral index (and its
error) and the magnetic field strength will also vary with position.
It is also not clear whether the thermal/non-thermal emission ratios follow the expectations of
the calibrations of \cite{mur11}, point-by-point.  
Nevertheless, the approach in which
the thermal/non-thermal ratio is tied to a SFR calibration, should be
an improvement upon previous estimates for other galaxies which usually assume a constant and 
singular thermal/non-thermal ratio for
every point in the galaxy.  These preliminary calculations, then,
suggest a 
total magnetic field strength of order 10 $\mu$G at the two sample positions selected.

\subsubsection{Rotation Measures and the Regular Field}
\label{sec:rm}

\begin{figure}[h]
   \centering
   \includegraphics*[width=3.5in]{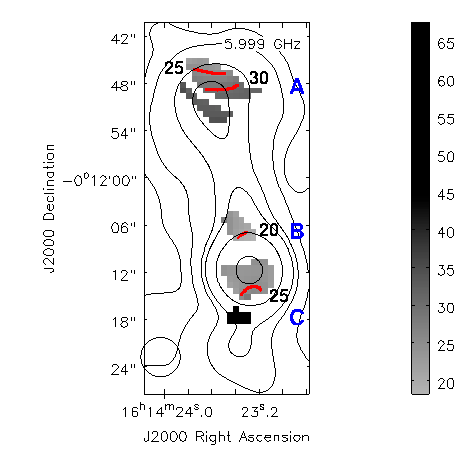}
   \caption{Rotation measure map, made from the L-band B-array and the
C-band C-array data,  at 5 arcsec resolution, computed at all pixels where the
polarized intensities in both bands exceed twice the rms noise. Several contours
(red) are marked, with all values  
 in units
of rad m$^{-2}$.   
The two regions with no contours marked have average values of
 33 rad m$^{-2}$ (northern region) and
62 rad m$^{-2}$ (southern). Letters denote the regions discussed in 
Sect.~\ref{sec:rm}.
 Black contours represent the 5 arcsec C-band C-array 
total intensity
image at 6.2, 30, 150, 300, 1200, and 6000 $\mu$Jy beam$^{-1}$. The beam is marked at lower left.
}
\label{fig:rot-measure}
\end{figure}

A powerful tool, which is potentially available for UGC~10288, is the technique of 
using 
the background radio galaxy, {\it CHANG-ES A},
 to probe the foreground magnetic field.
To this end, we computed the classical
Faraday rotation measure (RM) map  
(Fig.~\ref{fig:rot-measure}) between the
 B-array L-band and 
the C-array C-band data, at 5 arcsec resolution.
The
maximum $1\sigma$ error is $\pm\,10$ rad\,m$^{-2}$ at the outer edge of 
each patch, while the $1\sigma$ error of the mean RM of 
Fig.~\ref{fig:rot-measure} is 
$\pm\,2$ rad\,m$^{-2}$. The RM ambiguity is $\pm\,93$ rad\,m$^{-2}$.

The observed RM values are a superposition from 
{\it CHANG-ES A}, the galaxy UGC~10288, and the foreground of the Milky Way. The 
lobes of radio galaxies can have high RM values with a large dispersion, 
so that, when observed with low resolution, the average RM is small. The 
central sources of radio galaxies (jets) generally do not reveal 
significant intrinsic RM. The RM of {\it CHANG-ES A}'s northern lobe (at about 7~kpc 
projected distance above the plane of UGC~10288, marked `A' in
Fig.~\ref{fig:rot-measure}) of 
$+30\,\pm\,4$ rad\,m$^{-2}$ (average of the two visible regions)
agrees well with the foreground RM of the Milky 
Way \citep{tay09,opp12} 
at the Galactic coordinates of our radio maps (l=+12$^\circ$, 
b=+34$^\circ$). This indicates that no correction for the RM ambiguity is 
needed.

{The southernmost blob in Fig.~\ref{fig:rot-measure}
 (marked `C') is located behind the disk of 
UGC~10288, about 1.5~kpc above the plane, with an observed average value of
RM$\simeq+62\,\pm\,4$ rad\,m$^{-2}$.  We}
 interpret the (foreground 
subtracted) value of  RM$_i\simeq+32$ rad\,m$^{-2}$ as indication of a 
regular magnetic field in the galaxy disk, pointing towards us.
 A line-of-sight component of 
about +2~$\mu$G, a thermal electron density of 0.02~cm$^{-3}$ and a 
path length of 1~kpc 
can provide the observed RM$_i$. 
 If the path length is 
more than 1~kpc in the halo, the field strength and/or the electron
density lowers by the same 
factor.
However, as noted
earlier (Sect.~\ref{sec:overlays}), 
when high latitude radio emission is observed, the
features have generally been discrete with typical sizes of order
a kpc. In addition, disk exponential scale heights are also typically
$\approx$ 1 kpc.
Consequently, at the location of `C' at 1.5 kpc above mid-plane, we
might expect the effective path length of the magnetic field
to be approximately
that of a typical high latitude discrete feature.

RM$_i\simeq-7$ rad\,m$^{-2}$ {(observed average value of 
RM$\simeq+23\,\pm\,2$ rad\,m$^{-2}$ for both parts)}
towards the central region of the radio 
galaxy, at about 3.5~kpc above the plane of UGC~10288
(`B'), indicates that the 
field direction is reversed with respect to the plane. A thermal density 
in the halo of 0.01~cm$^{-3}$ requires a regular halo field with a 
line-of-sight component of about -1~$\mu$G, pointing away from us. 
Again, if the path length is 
more than 1~kpc in the halo, the field strength and/or the electron
density lowers by the same 
factor.  Between disk and halo, no regular field from the galaxy is 
detected, so that the RM is solely due to the Galactic foreground.

{The conclusion that a field reversal is present depends, of course, on the
accuracy of the RM measurements, especially region `C' which is
present in the total power maps only as diffuse emission between the core of
{\it CHANG-ES A} and its southern radio lobe.
If we
use a higher cut-off of 3$\sigma$ to form the RM map, then the value of `C'
becomes 64 rad m$^{-1}$, rather than the 62 rad m$^{-2}$ seen in Fig.~\ref{fig:rot-measure},
i.e. within the quoted  1$\sigma$ uncertainty of  $\pm$ 10 rad m$^{-2}$ on the RM at
the map edges (see above).
A  3$\sigma$ cut-off also reduces the  1$\sigma$ uncertainty in RM 
to  $\pm$ 7 rad m$^{-2}$;  however, such accuracy is not
required for this conclusion.  
}

A reversal between the azimuthal components of the disk field and halo 
field may be quite common among galaxies; it has been found from 
modeling the magnetic field in the Milky Way from extragalactic RMs
\citep[e.g.][]{sun08,jan12}
  and also in the northern part of the galaxy M51
\citep{fle11}. 
Dynamo models can hardly explain such reversals 
if the same dynamo process operates in disk and halo, unless wind 
outflows allow such ``mixed parities'' \citep{mos10}.


These results are necessarily preliminary, but
the fact that many spectral channels are available in each of our observations
(Table~\ref{table:obs}) lends itself well to the application of
the RM Synthesis technique \citep{bre05}
to UGC~10288.  This requires that (RA - DEC - Frequency) cubes be formed for all data sets,
and is an approach that
 will be attempted in the future.

\section{Discussion}
\label{sec:discussion}

Our observations have revealed some new surprises in
UGC~10288 and
the multi-array, multi-frequency approach has been essential in 
 understanding this galaxy and disentangling its emission from
the background radio galaxy.  Not only have we been able to 
correct for the emission of {\it CHANG-ES A}, but {\it CHANG-ES B}
would also have been assumed to be part of UGC~10288 
at low resolution (see 
Fig.~\ref{fig:Lband}a), had multiple arrays (i.e. higher resolutions) not been used.

Moreover, the combination of all arrays and both frequencies
has allowed us to detect high latitude
emission in UGC~10288.  However, rather than having
a broad scale halo,  
the high latitude features in UGC~10288 are
discrete and, in the case of the arc-like feature seen extending towards the
north on the eastern side of UGC~10288's disk (Fig.~\ref{fig:all_colour}b),
also large.  This feature extends to 21 arcsec ($~\sim$ 3.5 kpc)
above the disk. 

It is interesting to explore some properties of this arc.  For example,
taking the estimate of the total magnetic field
strength ($B\,\sim\,10$ $\mu$G, Sect.~\ref{sec:pol})  at the
position marked in Fig.~\ref{fig:overlays_a}b (13.5 arcsec, or 2.2 kpc above the plane)
and using standard formulae \citep[e.g.][]{hee09},
the particle lifetime to synchrotron losses is $t_{synch}\,\sim\,1.6\,\times\,10^7$ yr.
We do not know the outflow velocity, but it must be substantial in order to detect
synchrotron radiation at this $z$-height.  For example, if there were no additional
losses, this lifetime implies (for constant velocity) a vertical flow of
$v_z\,\sim\,134$ km s$^{-1}$, and higher still if adiabatic or other losses are
significant.
The largest uncertainty, apart from those introduced by assumptions,
belongs to $\alpha$ which affects $B$ by $\pm$ a few $\mu$G.
Since $t_{synch}\,\propto\,B^{-3/2}$, then the uncertainty on $v_z$ 
due to the uncertainty on $\alpha$ alone is $\sim\,$30\%.  What we see,
therefore, is that outflow speeds can be substantial in localized regions,
even in galaxies with modest SFRs.

As pointed out in Sect.~\ref{sec:overlays},
the disk exponential scale heights in regions outside of the discrete high-latitude
features
is less than 1 kpc (Sect.~\ref{sec:overlays}, 
Table~\ref{table:scaleheights}) and, even when one considers the scale height averaged
over regions with and without radio extensions, the average is only $\approx$ 1 kpc. 
This lack of a global radio halo, 
we suggest, is consistent with a rather modest SFR in UGC~10288
which we have modified downwards to be
0.4 - 0.5 M$_\odot$ yr$^{-1}$.  That is, discrete extensions 
may be formed in relation to underlying disk
activity, but not globally.  A comparison could be made to M~31 which has a similar
(slightly lower)
SFR of 0.3 M$_\odot$ yr$^{-1}$ \citep{ars11} and for which the synchrotron 
exponential scale height is only $\sim$ 300 pc \citep{fle04}.

It is clear that galaxies with similar global SFRs can be quite different in
other ways.  For example, although the SFRs of M~31 and M~33 are similar, their
star formation efficiencies, SFE = SFR/M(H$_2$), where 
M(H$_2$) is the molecular gas mass, differ by at least a factor of 3
\citep[][their Figs.~17 and 11, respectively]{tab10,gar07}\footnote{No molecular gas measurements
have yet been made for UGC~10288.}. 
Similarly, galaxies with comparably low SFRs might show differences in
the presence or absence of discrete
high latitude radio continuum features.  Possible reasons might be the degree of
clustering of star forming regions, or the relative strength and direction
of the magnetic fields.

When the various scale heights are averaged, irrespective of location,
we find beam-corrected scale heights of 
$h_e({\rm 4.1~GHz})\,\sim\,1.0$ kpc, 
$h_e({n_e})\,\sim\,1.2$ kpc, where we have doubled the observed average
H$\alpha$ scale height to take into account the fact that the observed
emission measure $\propto$ $n_e^2$, and
$h_e({12~\mu{\rm m}})\,\sim\,1.2$ kpc.  Consequently the warm dust and thermal
electron density scale heights are comparable and, arguably, so is the radio
scale height, although the latter is far more irregular
as Fig.~\ref{fig:overlays_a}b (from which the values of 
Table~\ref{table:scaleheights} were obtained) readily shows.

The high thermal fraction in UGC~1028 (44\% in C-band, Sect.~\ref{sec:sfr})
also has been seen before, again in galaxies with low or modest SFRs.
For example, \cite{dum00} find a C-band thermal fraction as high as 
54\% for 
the galaxy, NGC~5907, 
and \cite{chy07}
find thermal fractions of 50\% to 70\% in three late-type galaxies
at 4.85 GHz
compared to what is normally measured ($\approx$ 20\%).
These authors suggest that, in galaxies with higher SFRs, the non-thermal
fraction increases because of the non-linear dependence of non-thermal
luminosity on SFR, as noted in Sect.~\ref{sec:sfr}.  Our results depend
on an assumption of a linear dependence, but as already noted, we do
not anticipate a significant change at the low SFR of UGC~10288 unless
the non-thermal calibration is also markedly different.  Consequently,
our results of a high thermal fraction are consistent with previous
estimates for galaxies with similar SFRs.

A significant new result is, of course, the detection of 
a double-lobed radio galaxy, {\it CHANG-ES A}, behind the
disk of UGC~10288, which has allowed us to make some preliminary estimates
of the regular magnetic field in UGC~10288 prior to a full Rotation Measure
synthesis analysis.
Aside from the Milky Way, there have been only a few limited examples in which
polarized background sources have been used to probe foreground galaxies
\citep{han98,mao12} and these were for galaxies of large angular size (M~31
and the Large Magellanic Cloud).
In the case of UGC~10288, which is only 4.9 arcmin in diameter, the 90 degree alignment
of the double-lobed radio galaxy
is so fortuitous that both the disk and halo regions of the foreground galaxy can be probed.

For an adopted electron density of
$n_e\,=\,0.02$
cm$^{-3}$ a projected distance 1.5 kpc above the plane 
(position `C' in Fig.~\ref{fig:rot-measure})
and a path length of 1 kpc,
we find a regular field of $+$ 2 $\mu$G parallel to the line of sight.  
At 3.5 kpc above the plane (position `B' in Fig.~\ref{fig:rot-measure})
with $n_e\,=\,0.01$ cm$^{-3}$, $B_{||}\,\sim\,-1$ $\mu$G, suggesting a
reversal of the azimuthal field above the plane.

These values are still preliminary.  Nevertheless,
{\it CHANG-ES A} has the potential to be a powerful probe of
both  $B_{||}$ and $n_e$
in regions in which neither are measured directly via radio emission.
At the lower position (`C'), for example, the H$\alpha$ emission
has already dropped to the level of the noise (Fig.~\ref{fig:overlays_a}a).
This map has a
3$\sigma$ emission measure noise limit of 
13 pc cm$^{-6}$ \citep{ran96}
 which implies that any electron density less than
$n_e\,=\,0.1$ cm$^{-3}$ over the same path
length would not have been detected (adjusting as $1/\sqrt{l}$ for a different
path length, $l$).  

At the same time, our sampling of the total magnetic field strength in
the discrete northern arc, 2.5 kpc above the plane, reveals a rather
strong total magnetic field $B\,\sim\,10$ $\mu$G (Sect.~\ref{sec:pol}).
Although not at the same position as rotation measures were measured,
the implication is that field strengths in discrete high latitude features could 
be similar to those in the disk, possibly because of field compression 
in the arc.  
A similar result was found for
NGC~5775 \citep{soi11} for which X-shaped extensions have fields almost
as strong as in the disk, although the latter galaxy has, by contrast, a high SFR
and global radio halo.  NGC~4631 also has a relatively strong halo field
\citep{hum91a}.
It is unlikely that discrete strong high latitude features could be a result of
a mean-field dynamo alone.   Localized SF
activity in the underlying disk is likely contributing to this feature.

\section{Conclusions}
\label{sec:conclusions}

We have obtained JVLA wide-band radio continuum data of UGC~10288 in both total power as well as polarized intensity 
as part of the CHANG-ES program (Paper 1).  Three 
different array configurations (B, C, and D) at two frequencies, 1.5 GHz (L-band) and 6 GHz
(C-band) have been used and careful attention has been paid to the techniques needed to
reduce wide-band data.   Moreover, we have combined all total intensity data to
form a single map along with its spectral index, applicable to the intermediate frequency of 4.1 GHz.
The combined data image has revealed new features in UGC~10288 that were too faint to be seen in
the individual data sets.  

Our main conclusions are as follows:

$\bullet$  We have discovered a background double-lobed extragalactic radio source,
which we have named {\it {\it CHANG-ES A}} ($z_{phot}\,=\,0.39$)
 behind and perpendicular to the disk of UGC~10288.  
In previous observations, this
source had been blended with UGC~10288.  Approximately 50 arcsec to the west of this source, is another
background source, {\it {\it CHANG-ES B}} at the same redshift.

$\bullet$  We have disentangled the flux densities of UGC~10288 and {\it CHANG-ES A} and find, for
the foreground galaxy, that it is only 
 17\% and 11\% of the total blended flux density  
at L-band and C-band, respectively (Table~\ref{table:derived-ugc10288}).
UGC~10288 would not have been in the CHANG-ES survey had it not been blended with {\it CHANG-ES A}, since
its flux density is a factor of 5 below the minimum cut-off for the survey.

$\bullet$ We have revised the SFR of UGC~10288 downwards from the IRAS value of 1.3 M$_\odot$ yr$^{-1}$
to our new value of 0.4 - 0.5  M$_\odot$ yr$^{-1}$.  In the process, we have employed the SFR calibrations
of \cite{mur11} for both thermal and non-thermal radio emission emission, which also allows us to solve
for the global non-thermal spectral index ($\alpha_{NT}\,=\,-1.0$).

$\bullet$ The global thermal fractions tend to be high, as has been observed before in
galaxies which have lower SFRs, i.e. ${(S_{T}/S)}_{1.5}\,=\,0.17$ and
${(S_{T}/S)}_{6.0}\,=\,0.44$, at 1.5 GHz and 6 GHz, respectively.

$\bullet$  UGC~10288 shows discrete high latitude radio continuum features.  In particular, a large radio
continuum arc-like feature is observed to the north of the major axis on the east side of the galaxy
(Fig.~\ref{fig:all_colour}b).  This arc extends to 3.5 kpc above the plane.

$\bullet$ 
 At high resolution,
there is a trend for high latitude radio emission to occur above gaps in the 
underlying H$\alpha$ emission; however, we cannot confirm whether this relationship is a
physical one because of uncertainty in the line-of-sight location of the features.

$\bullet$ UGC~10288's radio continuum emission does not form a disk-wide global halo.
The beam-corrected exponential scale heights are typically $\sim$ 1 kpc when averaged over
 regions both with and without discrete
high latitude radio continuum features (smaller in regions which do not show such features).
High latitude radio emission therefore appears to occur in
localized discrete features in this galaxy.

$\bullet$ The fortuitous placement of {\it CHANG-ES A} has allowed us to do a preliminary rotation
measure analysis of UGC~10288 at several positions.  Between 1.5 and 3.5 kpc above mid-plane,
the azimuthal components of magnetic field appear to reverse directions.  

$\bullet$ The minimum energy total
magnetic field strength in the large north-eastern arc at a z = 2.5 kpc is $\sim$ 10 $\mu$G
at a location 2.5 kpc above the plane, indicating that the magnetic field is substantial in
this high latitude feature.

\clearpage

\acknowledgments
We thank Jeremy MacHattie for retrieving previous radio flux density information at the
time of proposal submission. JAI would like to thank the Natural Sciences and Engineering
Research Council of Canada for grant.
 This research has used the Karl G. Jansky Very Large Array operated
by the National Radio Astronomy Observatory (NRAO).  NRAO
 is a facility of the
National Science Foundation operated under cooperative
agreement by Associated Universities, Inc.
This publication makes use of data products from the Wide-field Infrared Survey Explorer, 
which is a joint project of the University of California, Los Angeles, and the Jet Propulsion 
Laboratory/California Institute of Technology, funded by the National Aeronautics and 
Space Administration. 
This research has made use of the NASA/IPAC Extragalactic Database (NED) which is operated by the 
Jet Propulsion Laboratory, California Institute of Technology, under contract with the National 
Aeronautics and Space Administration. 
This work is based in part on observations made with the Spitzer Space Telescope, 
which is operated by the Jet Propulsion Laboratory, California Institute of Technology under a contract 
with NASA.
Funding for SDSS-III has been provided by the Alfred P. Sloan 
Foundation, the Participating Institutions, the National Science 
Foundation, and the U.S. Department of Energy Office of Science. The 
SDSS-III web site is http://www.sdss3.org/.




{\it Facilities:} \facility{JVLA} 

\clearpage



\clearpage

\clearpage
\begin{deluxetable}{lc}
\tabletypesize{\scriptsize}
\renewcommand{\arraystretch}{0.90}
\tablecaption{Properties of UGC~10288$^a$\label{table:prop-ugc10288}}
\tablewidth{0pt}
\tablehead{
{Property}  & Value\\
}
\startdata
Right Ascension (J2000)                         &  16h14m24.80s \\
Declination  (J2000)                            & -00d12m27.1s \\
Inclination (deg)                               & 90\\
d$_{25}$ (arcmin)\tablenotemark{b}              & 4.9\\
$V_\odot$ (km/s)\tablenotemark{c}               & 2046\\
D (Mpc)\tablenotemark{d}                        & 34.1\\ 
Morphological Type                              & Sc\\
S$_{1.4~{\rm GHz}}$ (mJy)\tablenotemark{e}      & 26.1\\
$L_{FIR}$ ($L_\odot$)\tablenotemark{f}          & 2.55$\,\times\,10^9$\\
SFR$_{IRAS}$ (M$_\odot$ yr$^{-1}$)\tablenotemark{g}      & 1.3\\
SFR$_{this~work}$ (M$_\odot$ yr$^{-1}$)\tablenotemark{h}  & 0.4 - 0.5\\
$\rho$ (Mpc$^{-3}$)\tablenotemark{i}            & 0.23\\
W$_{20}$ (km s$^{-1}$)\tablenotemark{j}          & 397.3\\
M$_T$ (M$_\odot$)\tablenotemark{k}               & 2.2$\,\times\,10^{11}$\\
S$_{int}$ (Jy km s$^{-1}$)\tablenotemark{l}      &36.4\\
M$_{HI}$ (M$_\odot$)\tablenotemark{m}            & 9.9$\,\times\,10^9$ \\  
\enddata
\tablenotetext{a}{See Paper 1 for original sources if not specified.}
\tablenotetext{b}{Observed blue diameter at the 25th mag/arcsec$^2$ isophote.}
\tablenotetext{c}{Heliocentric systemic velocity.}
\tablenotetext{d}{Distance, assuming H$_O$ = 73 km s$^{-1}$ Mpc$^{-1}$ and correcting for 
Virgo Cluster and Great Attractor perturbations.}
\tablenotetext{e}{Previously measured flux density at a frequency of 1.4 GHz.}
\tablenotetext{f}{Far infrared luminosity.}
\tablenotetext{g}{Star formation rate based on IRAS photometry.}
\tablenotetext{h}{See Sect.~\ref{sec:sfr} for an explanation.}
\tablenotetext{i}{Density of galaxies brighter than -16 mag in the vicinity of UGC~10288.}
\tablenotetext{j}{Width of the HI profile at the 20\% level from the HI Parkes All Sky Survey (HIPASS) Catalogue \citep{mey04}.}
\tablenotetext{k}{Total mass, using M$_T\,=\,{\rm W}_{20}^2\,{\rm d}_{25}/(8\,G)$ (cgs units). Note that this result is
a factor of 2 higher than the quantity listed in Paper I which was taken from older HI data.}
\tablenotetext{l}{HIPASS integrated flux density of the HI line \citep{mey04}.
This value may be compared with S$_{int}\,=\,38.2~{\rm Jy~ km\, s^{-1}}$
and S$_{int}\,=\,34.7~{\rm Jy~ km\, s^{-1}}$ using the NRAO 43 m and Green Bank Telescopes, 
respectively \citep{hog07}.}
\tablenotetext{m}{HI mass, from $[{\rm M}_{HI}/{\rm M_\odot}]\,=\,2.35\,\times\,10^5 [D/{\rm Mpc}]^2\,
[S_{int}/({\rm Jy~ km\, s^{-1})}]$. }
\end{deluxetable}

\clearpage

\begin{deluxetable}{lccccc}
\tabletypesize{\scriptsize}
\renewcommand{\arraystretch}{0.90}
\tablecaption{UGC~10288 Observing and Calibration Information$^a$\label{table:obs}}
\tablewidth{0pt}
\tablehead{
{Frequency}                   & \multicolumn{3}{c}{1.5 GHz (L band)}  & 
\multicolumn{2}{c}{6.0 GHz (C band)} \\
{Array} & B & C & D & C & D 
}
\startdata
Date of Observations                               & 5 Apr. 2011 & 30 Mar. 2012& 30 Dec. 2011 & 11 Feb. 2012& 10 Dec. 2011 \\
                                                   &             &             &              & 16 Feb. 2012&              \\
Frequency range (GHz)\tablenotemark{b}             &1.247$\to$1.503 &1.247$\to$1.503 & 1.247$\to$1.503& 4.979$\to$7.021
                                                                         &4.979$\to$7.021   \\
                                                   &1.647$\to$ 1.903 &1.647$\to$ 1.903 &1.647$\to$ 1.903 & &  \\
Total bandwidth (MHz)                              & 512 &512 &512 & 2042 & 2042   \\
No. of spectral windows                            & 32  & 32 & 32 & 16 & 16\\
No. of channels per spectral window                & 64  & 64 & 64 & 64 & 64\\
Total no. channels                                 & 2048 & 2048 & 2048 & 1024 & 1024 \\
Channel separation (MHz)                           & 0.25 &0.25 &0.25 & 2.0& 2.0   \\
Spectral resolution (MHz)\tablenotemark{c}         & 0.50 &0.50 &0.50 & 4.0& 4.0  \\
Integration time (s)\tablenotemark{d}              & 10 & 10 & 10 & 10 & 10  \\ 
Obs. Time (min)\tablenotemark{e}                   & 96.7 & 43.0 & 18.3&191 & 38.7   \\
Primary calibrator\tablenotemark{f}                   & 3C286 &3C286 &3C286 & 3C286& 3C286  \\
Secondary calibrator\tablenotemark{g}                  & J1557--0001& J1557--0001&J1557--0001 &J1557--0001 &J1557--0001 \\
~~~~~$S_{\nu_0}$(Jy)\tablenotemark{h}   &0.487 $\pm$ 0.004 & 0.461 $\pm$ 0.002 & 0.47 $\pm$ 0.01 &0.429 $\pm$ 0.001 & 
0.432 $\pm$ 0.002 \\
                                                   & &                         &    &0.422 $\pm$ 0.004& \\
Pol.-leakage calibrator\tablenotemark{i}           & J1407+2827 &J1407+2827 &J1407+2827 &J1407+2827 & J1407+2827  \\
~~~~~$S_{\nu_0}$(Jy)\tablenotemark{h}  &0.953 $\pm$ 0.004 &0.963 $\pm$ 0.003 &0.943 $\pm$ 0.003 & 2.094 $\pm$ 0.002 
& 2.103 $\pm$ 0.002\\
\enddata
\tablenotetext{a}{These values apply to the set-up of the observations and do not take into account the flagging
of bad data.}
\tablenotetext{b}{The frequency range of 1.503 GHz to 1.647 GHz was avoided due to interference. At C-band,
the upper and lower portions of the band overlap slightly at the center; consequently, 
the total bandwidth is slightly less than 1024
channels $\times$ 2 MHz/channel = 2048 MHz.}
\tablenotetext{c}{After Hanning smoothing.}
\tablenotetext{d}{Single record measurement time.}
\tablenotetext{e}{On-source observing time before flagging.}
\tablenotetext{f}{This source was also used as the bandpass calibrator and for determining the
absolute position angle for polarization.}
\tablenotetext{g}{This source is a `primary' calibrator in the sense of its amplitude errors 
($<$ 3\% amplitude closure errors expected) 
in all arrays and in both bands. It is separated from UGC~10288 by 4.14 degrees in the sky.} 
\tablenotetext{h}{Flux density of the calibrator specified on the previous line, at a representative frequency, $\nu_0$, within
the frequency band.  For L-band, we have taken $\nu_0\,=\,$1.495 GHz and for C-band,  $\nu_0\,=\,$ 5.94 GHz.  Uncertainties are based on
the extrapolation of amplitudes and phases from 3C286 after phase, amplitude, and bandpass calibrations have been applied.  
Uncertainties at other frequencies
within the respective bands are similar.
For the C-array C-band
data, the two rows correspond to the two dates of these observations, in the same order as the listed dates.}
\tablenotetext{i}{This zero-polarization calibrator, also known as OQ208 and QSO B1404+2841, was used to
determine the polarization leakage terms.}

\end{deluxetable}

\clearpage

\begin{deluxetable}{lccccc}
\tabletypesize{\scriptsize}
\tablecaption{UGC~10288 Map Parameters (Individual Data Sets)\label{table:map-ugc10288}}
\tablewidth{0pt}
\tablehead{
{Frequency}                   & \multicolumn{3}{c}{1.5 GHz (L band)}  & 
\multicolumn{2}{c}{6.0 GHz (C band)} \\
{Array} & B & C & D & C & D 
}
\startdata
uv weighting\tablenotemark{a} & Briggs & Briggs  & Briggs 
& Briggs & Briggs \\
No. Self-cals\tablenotemark{b} &  none    & 1 a\&p  & 1 a\&p 
 &   2 a\&p + 1 p      &   1 a\&p     \\
\hline
 & \multicolumn{5}{c}{I images}\\
\hline
Clean spatial scales (arcsec)\tablenotemark{c}  &0, 7.5, 15 & 0, 20, 40, 80 & 0 & 0, 5, 10, 20 & 0, 10, 20, 40, 60\\
Synth. beam\tablenotemark{d} \\
~~~~~~($^{\prime\prime}$, $^{\prime\prime}$, $^\circ$)  &3.80, 3.58, 66.2 & 12.18, 10.71, -55.0
&40.23, 34.34,  -31.3 & 3.14, 2.87, -11.0 & 10.96, 9.46, -28.5\\
Map peak(mJy beam$^{-1}$)\tablenotemark{e} & 9.57 & 12.6 & 19.2 & 7.62 & 8.94 \\
rms ($\mu$Jy beam$^{-1}$)\tablenotemark{f} & 14 & 22 & 42 & 3.1  & 7.0 \\
\hline
 & \multicolumn{5}{c}{Q \& U images\tablenotemark{g}}\\
\hline
Synth. beam\tablenotemark{h} & & & &  & \\
~~~~~~($^{\prime\prime}$, $^{\prime\prime}$, $^\circ$)  & 3.72, 3.54, 60.8
  & 12.18, 10.71, -55.0 & 40.19, 34.33, -31.4
 &  3.14, 2.87, -11.0 & 10.95, 9.47, -28.4\\
rms ($\mu$Jy beam$^{-1}$)\tablenotemark{i} & 13  & 22  & 35  & 2.9 &  5.4\\
\hline
 & \multicolumn{5}{c}{P images\tablenotemark{j}}\\
\hline
Map peak($\mu$Jy beam$^{-1}$)\tablenotemark{e} &  177.9 & 178.8 & 272.1 &  149.5 & 146.7 \\
\enddata
\tablecomments{These parameters represent
the images presented in Figs.~\ref{fig:Lband},  \ref{fig:Cband}, \ref{fig:Lband-pol} and \ref{fig:Cband-pol}.}
\tablenotetext{a}{See \citet{bri95} for a description of Briggs weighting with various
`robust' factors.  Robust = 0 was used in each case, as employed in the CASA {\it clean} task.}
\tablenotetext{b}{Number of self-calibration iterations, where 'p' refers to phase and `a\&p' refers to amplitude and
phase together.}
\tablenotetext{c}{Scales used for the multi-scale clean.  The scale, 0, corresponds to a classic clean for which the emission
is assumed to be described by point sources.}
\tablenotetext{d}{Synthesized beam FWHM of major and minor axis, and position angle.}
\tablenotetext{e}{Maximum value on the map.}
\tablenotetext{f}{Rms map noise before primary beam correction.}
\tablenotetext{g}{Stokes Q and U images.  The {\it clean} spatial scales used were equivalent to the I images with the
highest spatial scale dropped (except for D-array, L-band images for which only the zero spatial scale was used). }
\tablenotetext{h}{The cross-hands (RL, LR) had slightly different sets of flags applied to them compared to the
parallel hands (RR, LL), leading to small differences in the synthesized beams 
for Q and U compared to I.}
\tablenotetext{i}{The rms noise is lower than for I because the lower and less extensive polarized emission results in fewer
clean errors.}
\tablenotetext{j}{Polarized intensity images, where P = $\sqrt{Q^2\,+\,U^2}$, corrected for bias.}
\end{deluxetable}

\clearpage

\begin{deluxetable}{lcccc}
\tabletypesize{\scriptsize}
\renewcommand{\arraystretch}{0.90}
\tablecaption{UGC~10288 Map Parameters (Combined Data Sets)\label{table:combined}}
\tablewidth{0pt}
\tablehead{
{Parameter}   & BCD Lband & CD Cband & BCD+CL (rob0) & BCD+CL (uvtap) \\
}
\startdata
Central frequency (GHz)\tablenotemark{a}  & 1.575 & 5.998 &4.134& 4.134\\
uv weighting\tablenotemark{b} & Briggs + 10 k$\lambda$ taper  & Briggs + 16 k$\lambda$ taper  & Briggs & Briggs + 16 k$\lambda$ taper \\
Clean spatial scales (arcsec) & 0, 5, 10 & 0, 5, 10, 20 & 0, 7.5, 15, 30\tablenotemark{c} &0, 15, 30, 60\\
Synth. beam \\
~~~~~~($^{\prime\prime}$, $^{\prime\prime}$, $^\circ$)  & 11.43, 10.11, -84.23 & 6.30, 6.01, -81.30 & 3.77, 3.40, -3.89  & 6.81, 6.46, -82.97\\
Map peak(mJy beam$^{-1}$) & 12.06 & 8.45 & 8.41 & 9.43 \\
rms ($\mu$Jy beam$^{-1}$) & 20 & 4.0 & 7.0 & 10\\
\enddata
\tablecomments{These parameters represent images shown in Figs.~\ref{fig:all_colour} to ~\ref{fig:overlays_b}.
The meaning of each parameter is as specified in 
Table~\ref{table:map-ugc10288} unless otherwise indicated.}
\tablenotetext{a}{Frequency of the maps, intermediate between L and C-bands.}
\tablenotetext{b}{Robust = 0 was used in each case. The term, `uvtap', 
means that a uv taper, of specified scale, was applied to the data.}
\tablenotetext{c}{The largest spatial scale was dropped when this clean was very advanced.}
\end{deluxetable}

\clearpage

\begin{deluxetable}{lcc}
\tabletypesize{\scriptsize}
\renewcommand{\arraystretch}{0.90}
\tablecaption{Derived Properties of UGC~10288 and {\it CHANG-ES A}\label{table:derived-ugc10288}}
\tablewidth{0pt}
\tablehead{
{Property}   & UGC~10288 & {\it CHANG-ES A} \\
}
\startdata
Right Ascension (J2000)\tablenotemark{a}  &   16h14m24.80s & 16h14m23.28s $\pm$ 0.04s \\
Declination  (J2000))\tablenotemark{a}    & -00d12m27.1s &  -00d12m11.6 $\pm$ 0.5s  \\
$S_L$ (mJy)\tablenotemark{b}   & 4.4 $\pm$ 0.5    &  22 $\pm$ 2            \\
$S_C$ (mJy)\tablenotemark{b}    &  1.53 $\pm$ 0.05      &  12.0 $\pm$ 0.2    \\
$\alpha_{\rm total}$\tablenotemark{c}   & -0.76 $\pm$ 0.11 &  -0.44 $\pm$ 0.08     \\
\enddata
\tablenotetext{a}{U10288 position from NED.  For {\it CHANG-ES A}, the position
 has been measured from the highest resolution data.}
\tablenotetext{b}{Flux densities in L or C bands,
measured from the primary beam-corrected images as described in Sect.~\ref{sec:results}.  Uncertainties
represent variations from adjusting the box size containing the emission.}
\tablenotetext{c}{Global spectral index accordinging to $S_{L}/S_{C} = ({\nu_L/\nu_C})^\alpha$, where
$\nu_L$ and $\nu_C$ are the L-band and C-band central frequencies, respectively.}
\end{deluxetable}

\clearpage
\begin{deluxetable}{lcccccc}
\tabletypesize{\scriptsize}
\renewcommand{\arraystretch}{0.90}
\tablecaption{Point-like Sources with Radio Emission near UGC~10288\label{table:pointsources}}
\tablewidth{0pt}
\tablehead{
{Source}   & RA (J2000)$^a$ & DEC (J2000)$^a$ & SDSS Identifier$^b$ 
& $S_L$$\,^c$ & $S_C$$\,^c$ & $\alpha$$\,^d$\\
              & (h m s)        &  $^\circ$ $^\prime$ $^\prime\prime$ & & ($\mu$Jy)& ($\mu$Jy)\\
}
\startdata
1 &   16 14 29.88  & -00 12 01.2 & J161429.87-001201.3 (S) & 50 $\pm$ 10 & 9  $\pm$ 4 & -1.3  $\pm$ 0.6\\
2 &   16 14 28.80  & -00 12 13.6 & ND & $<$ 16 & 16 $\pm$ 4 & $>$ 0\\
3 &   16 14 27.23  & -00 13 21.0 & ND & 35 $\pm$ 8 & 17  $\pm$ 4 & -0.5 $\pm$ 0.4 \\
4 &   16 14 25.43  & -00 12 12.1 & J161425.40-001212.2 (S) & $<$ 16 & 16 $\pm$ 4 & $>$ 0\\
5 &   16 14 24.87  & -00 13 18.5 & J161424.84-001318.3 (S) & 109 $\pm$ 5 & 53 $\pm$ 4 & -0.5 $\pm$ 0.1\\
6 &   16 14 21.40  & -00 12 36.6 & ND & 43 $\pm$ 8 & 17 $\pm$ 4 & -0.7 $\pm$ 0.3\\
CHANG-ES B &   16 14 19.80  & -00 11 55.6 &  J161419.79-001155.6 (G) & 3290 $\pm$ 50 &1040 $\pm$ 20 & -0.83 $\pm$ 0.02  \\
7 &   16 14 19.04  & -00 11 34.4 & ND & 28 $\pm$ 6 & 12 $\pm$ 4 & -0.634 $\pm$ 0.4 \\
\enddata
\tablenotetext{a}{Measured from the $\lambda\,$3.6 $\mu$m map at the maximum value (Fig.~\ref{fig:spitzer}).
The uncertainty is approximately 0.8 arcsec (one-half of the spatial resolution).
}
\tablenotetext{b}{SDSS identification number.  `ND' means no detection in the SDSS. (S) = `Star'; (G) =
`Galaxy'}
\tablenotetext{c}{C-band and L-band flux densities were measured from the primary-beam corrected 
combined C+D array images and the combined B+C array images, respectively.
Uncertainties represent a range resulting from varying the measurement box size.}
\tablenotetext{d}{Spectral index, determined from the previous two columns.}
\end{deluxetable}

\clearpage

\begin{deluxetable}{lccc}
\renewcommand{\arraystretch}{0.90}
\tablecaption{Disk Exponential Scale Heights\label{table:scaleheights}}
\tablewidth{0pt}
\tablehead{
Strip \# & BCD+CL (4.1 GHz) & H$\alpha$ & WISE $\lambda\,$12 $\mu$m \\
}
\startdata
1 (North) & 7.9 (1.3) & 2.3 (0.38) & 6.4 (1.1)\\
~~$\,$(South) & 5.1 (0.85) & 5.1 (0.84) & 7.7 (1.3)\\
2 (North) & 5.2 (0.86)& 3.2 (0.54)& 6.8 (1.1)\\
~~$\,$(South)  & 2.8 (0.46) & 4.0 (0.66) & 6.8 (1.1)\\
4 (North) & 8.6 (1.4) & 2.9 (0.47)& 7.5 (1.2)\\
~~$\,$(South) & 6.7 (1.1)\tablenotemark{a} &4.0 (0.66) & 7.1 (1.2)\\
\enddata
\tablecomments{Notes. Comparison of {\it beam corrected} exponential scale heights between the radio 
emission from combined array,
combined frequency data (4.1 GHz), the H$\alpha$ emission measure map, and the WISE $\lambda\,$12 $\mu$m map,
all at 7 arcsec
resolution.  Corrections for non-zero baselines have been made. Values are in arcsec with kpc in parentheses.}
\tablenotetext{a}{This value reduces to 4.6 (0.76) if the fit is cut at the first `shoulder' of emission.}
\end{deluxetable}





\end{document}